\documentclass[journal,onecolumn]{IEEEtran}
\usepackage{cite}

\ifCLASSINFOpdf
\else
\fi
\usepackage[cmex10]{amsmath}
\usepackage{acronym} 
\usepackage{psfrag}   
\usepackage{bbm}
\usepackage{multirow,booktabs}
\usepackage{graphicx,subfig}
\usepackage{framed}
\usepackage{algorithmic}
\usepackage{setspace}

\begin{document}
%
\title{Multi-Target Tracking in Distributed Sensor Networks using Particle {PHD} Filters}
%
%
%

\author{Mark~R.~Leonard,~\IEEEmembership{Student Member,~EURASIP,} 
        and~Abdelhak~M.~Zoubir,~\IEEEmembership{Member,~EURASIP,}
\thanks{The authors are with the Signal Processing Group, Institute of Telecommunications, Technische Universit\"at Darmstadt, Darmstadt 64283, Germany (e-mail: leonard@spg.tu-darmstadt.de, zoubir@spg.tu-darmstadt.de)}
\thanks{Manuscript submitted November 27, 2018}
}

%
%

\markboth{Submitted for publication}%
{}
%



\maketitle


\newacro{ac}[AC]{aggregation chain}
\newacro{ac-dpf}[AC-DPF]{aggregation-chain distributed particle filter}
\newacro{atb}[ATB]{Adaptive Target Birth}
\newacro{atc}[ATC]{Adapt-then-Combine}
\newacro{av}[AV]{neighborhood averaging}
\newacro{avn}[AV-$n$]{neighborhood averaging every $n^\text{th}$ iteration}
\newacro{av-bc}[AV-BC]{neighborhood averaging with neighborhood broadcast}
\newacro{asn}[ASN]{autonomous sensor network}
\newacro{cl}[CL]{cooperative localization}
\newacro{cr}[CR]{coverage ratio}
\newacro{cw-dpf}[CW-DPF]{consensus-weight distributed particle filter}
\newacro{dkf}[DKF]{diffusion Kalman filter}
\newacro{dpf}[DPF]{diffusion particle filter}
\newacro{ddf-pphdf}[DDF-PPHDF]{Distributed Data Fusion Particle PHD Filter}
\newacro{evsm}[EVSM]{extended virtual spring mesh}
\newacro{fov}[FOV]{field of view}
\newacro{fovs}[FOVs]{fields of view}
\newacro{gps}[GPS]{global positioning system}
\newacro{ins}[INS]{inertial navigation system}
\newacro{jpdaf}[JPDAF]{Joint Probabilistic Data Association Filter}
\newacro{kf}[KF]{Kalman Filter}
\newacro{la}[LA]{leader agent}
\newacro{ler}[LER]{localization error ratio}
\newacro{lvp}[LVP]{localized Voronoi polygon}
\newacro{map}[MAP]{maximum a posteriori}
\newacro{mht}[MHT]{Multiple Hypothesis Tracker}
\newacro{mmse}[MMSE]{minimum mean-square-error}
\newacro{mtt}[MTT]{multi-target tracking}
\newacro{ncon}[NCON]{neighborhood consensus}
\newacro{nnsf}[NNSF]{nearest neighbor standard filter}
\newacro{ospa}[OSPA]{Optimal Subpattern Assignment}
\newacro{crlb}[CRLB]{Cram\'{e}r-Rao Lower Bound}
\newacro{pcrlb}[PCRLB]{Posterior Cram\'{e}r-Rao Lower Bound}
\newacro{dpcrlb}[DPCRLB]{Distributed Posterior Cram\'{e}r-Rao Lower Bound}
\newacro{pdf}[PDF]{probability density function}
\newacro{pf}[PF]{Particle Filter}
\newacro{phd}[PHD]{Probability Hypothesis Density}
\newacro{phd-dpf}[D-PPHDF]{Diffusion Particle PHD Filter}
\newacro{phd-pf}[PPHDF]{Particle PHD Filter}
\newacro{ms-pphdf}[MS-PPHDF]{Multi-Sensor Particle PHD Filter}
\newacro{rfs}[RFS]{random finite set}
\newacro{rmse}[RMSE]{root-mean-square error}
\newacro{roi}[ROI]{region of interest}
\newacro{sir}[SIR]{sampling importance resampling}
\newacro{sis}[SIS]{sequential importance sampling}
\newacro{sla-dpf}[SLA-DPF]{single-leader--agent distributed particle filter}
\newacro{snr}[SNR]{signal-to-noise ratio}
\newacro{stt}[STT]{single-target tracking}
\newacro{vf}[VF]{virtual forces}
\newacro{vf-c}[VF+EVSM]{VF combined with EVSM}
\newacro{vf-nh}[VF-NH]{VF with acute-angle neighborhood selection}
\newacro{vf-nh-c}[VF-NH+EVSM]{VF-NH combined with EVSM}
\newacro{wsn}[WSN]{wireless sensor network}

\begin{abstract}
Multi-target tracking is an important problem in civilian and military applications. This paper investigates multi-target tracking in distributed sensor networks. Data association, which arises particularly in multi-object scenarios, can be tackled by various solutions. We consider sequential Monte Carlo implementations of the \ac{phd} filter based on random finite sets. This approach circumvents the data association issue by jointly estimating all targets in the region of interest. To this end, we develop the Diffusion Particle PHD Filter (D-PPHDF) as well as a centralized version, called the Multi-Sensor Particle PHD Filter (MS-PPHDF). Their performance is evaluated in terms of the Optimal Subpattern Assignment (OSPA) metric, benchmarked against a distributed extension of the Posterior Cram\'{e}r-Rao Lower Bound (PCRLB), and compared to the performance of an existing distributed PHD Particle Filter. Furthermore, the robustness of the proposed tracking algorithms against outliers and their performance with respect to different amounts of clutter is investigated.
\end{abstract}

\begin{IEEEkeywords}
Multi-target tracking, distributed target tracking, Particle Filter, PHD Filter, robustness, Posterior Cram\'{e}r-Rao Lower Bound 
\end{IEEEkeywords}

%
\IEEEpeerreviewmaketitle

\section{Introduction}
%
%
%
%

\IEEEPARstart{T}{he} problem of \ac{mtt} is becoming increasingly important in many military and civilian applications such as air and ground traffic control, harbor surveillance, maritime traffic control, or video communication and surveillance \cite{challa2011fundamentals, maresca2014maritime, rambach2015collaborative}. Distributed sensor networks offer a desirable platform for \ac{mtt} applications due to their low cost and ease of deployment, their lack of a single point of failure, as well as their inherent redundancy and fault-tolerance \cite{olfati-saber2007consensus}. A comprehensive overview of the state-of-the-art of distributed \ac{stt} is given in \cite{hlinka2013distributed}. Distributed versions of the Kalman Filter \cite{hlinka2013distributed, cattivelli2008diffusion} and its nonlinear, non-Gaussian counterpart, the \ac{pf} \cite{arulampalam2002a-tutorial}, have been well-studied. However, they cannot be applied directly to \ac{mtt} as they do not account for the problem of data association. Although there are methods such as the \ac{jpdaf} \cite{bar-shalom2011tracking} or the \ac{mht} \cite{reid1979an-algorithm} that address this problem in \ac{stt} algorithms, the resource constraints in sensor networks might pose a challenge on finding suitable distributed implementations \cite{oh2007tracking}. The \acl{phd} (\acs{phd}) filter \cite{clark2006multiple, mahler2003multitarget}, in contrast, resorts to the concept of \acl{rfs}s (\acs{rfs}s) to circumvent the problem of data association altogether. 

\begin{figure*}[!t]
\psfrag{Track 1}[bc][bc][0.7]{\footnotesize Target 1}
\psfrag{Track 2}[bc][bc][0.7]{\footnotesize Target 2}
\psfrag{Track 3}[bc][bc][0.7]{\footnotesize Target 3}
\psfrag{y}[tc][tc][0.9]{\footnotesize y[m]}
\psfrag{x}[tc][tc][0.9]{\footnotesize x[m]}
\psfrag{0}[Bc][Bc][0.7]{\footnotesize 0}
\psfrag{5}[cc][cc][0.7]{\footnotesize 5}
\psfrag{15}[cc][cc][0.7]{\footnotesize 15}
\psfrag{25}[cc][cc][0.7]{\footnotesize 25}
\psfrag{10}[cc][cc][0.7]{\footnotesize 10}
\psfrag{20}[cc][cc][0.7]{\footnotesize 20}
\psfrag{30}[cc][cc][0.7]{\footnotesize 30}
\psfrag{-10}[cc][cc][0.7]{\footnotesize -10}
\psfrag{-20}[cc][cc][0.7]{\footnotesize -20}
\psfrag{-30}[cc][cc][0.7]{\footnotesize -30}
\psfrag{-5}[cc][cc][0.7]{\footnotesize -5}
\psfrag{-15}[cc][cc][0.7]{\footnotesize -15}
\psfrag{-25}[cc][cc][0.7]{\footnotesize -25}
\centering
\subfloat[]{
\includegraphics[width=3in]{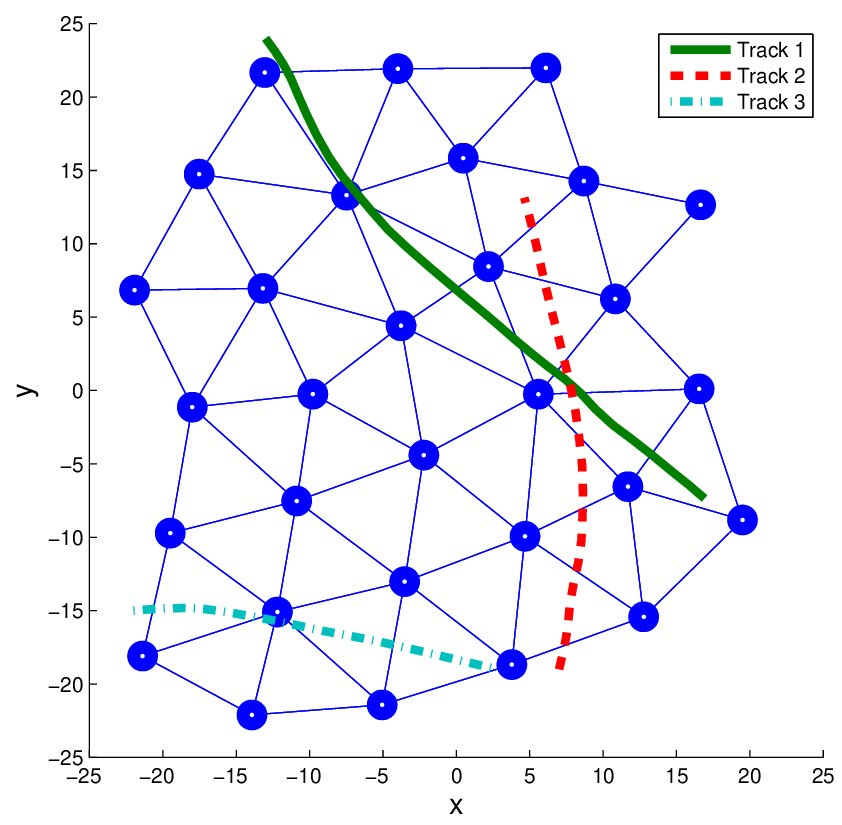}
\label{fig:mtt_tracks}
}\qquad
\subfloat[]{
\includegraphics[width=3in]{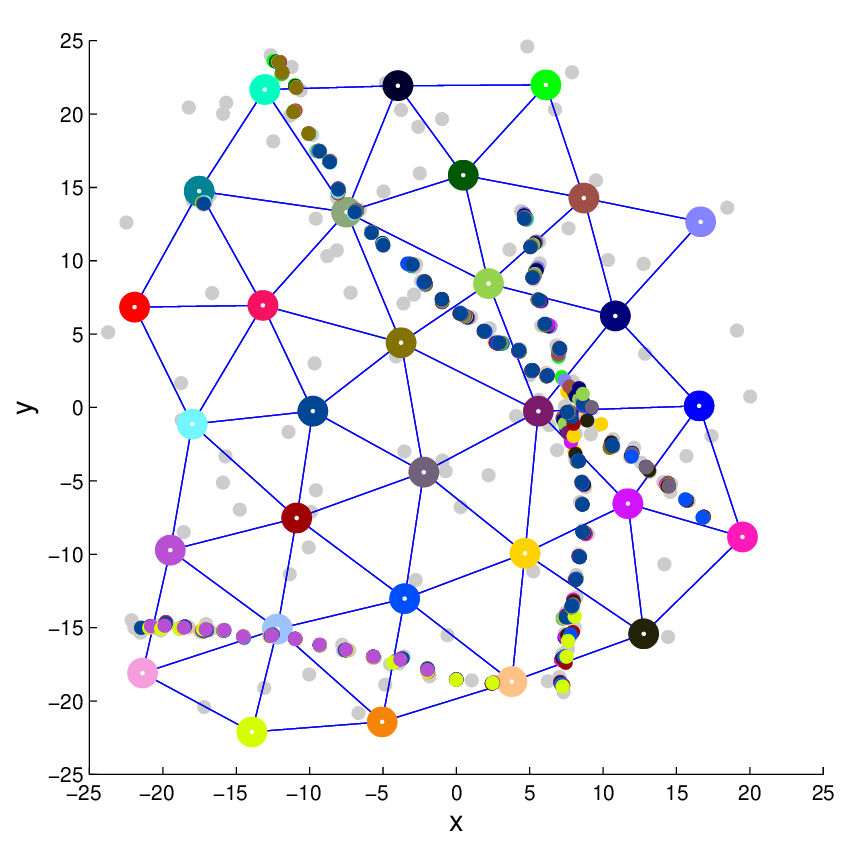}
\label{fig:subtracks}
}
\caption{(a) Distributed sensor network with 1-coverage of the region of interest and 3 exemplary target tracks. (b) Example of tracking 3 targets with the \acl{phd-dpf} (\acs{phd-dpf}). The small colored dots represent the target location estimates obtained by the respective node with the same color. The light gray dots show the collective measurements obtained by all nodes in the network.}

\end{figure*}

In this work, we investigate distributed \ac{mtt} in a sensor network with 1-coverage of the \ac{roi}, i.e., the sensor nodes have non- or barely overlapping \ac{fovs} and are distributed such that maximum area coverage is attained \cite{wang2003integrated}. An exemplary network layout with these properties is depicted in Figure \ref{fig:mtt_tracks}). Autonomous distribution algorithms for realizing such a topology have been studied in our previous work \cite{balthasar2014}. The nodes in the network communicate with their neighbors in order to collaboratively detect and track targets in the \ac{roi}. In addition, all of the sensors are equipped with a signal processing unit, allowing them to form decisions without a fusion center. That way, the network can autonomously react to events such as the detection of an intruder without relying on a network operator. For the sake of simplicity, the network is considered to be static. However, the consideration of mobile sensor nodes would enable reactions such as target pursuit or escape.

Since the \acs{fov} and communication radius of each node are limited, a target is only seen by a subset of the network, which changes as the target moves through the \ac{roi}. Hence, at each time instant, there is an \emph{active} and an \emph{inactive} part of the network. The goal, thus, is to detect and observe the target in a distributed and collaborative fashion as it travels across the \ac{roi}, rather than reaching a network-wide consensus on its state and have the estimate available at each node.


In the sequel, we develop a distributed Particle PHD filter called \ac{phd-dpf}, which uses neighborhood communication to collaboratively estimate and track a single-sensor \ac{phd} at each node in the active subnetwork. In addition, we formulate the \acl{ms-pphdf} (\acs{ms-pphdf}), a centralized extension of the \ac{phd-dpf}. The performance of both algorithms is evaluated in terms of the \acl{ospa} (\acs{ospa}) metric \cite{Schuhmacher2008}, which is calculated for the joint set of target state estimates of the active subnetwork. Furthermore, a distributed version of the \acl{pcrlb} (\acs{pcrlb}) \cite{van2004detection, Hue2002, Hue2002a}---again averaged over the active subnetwork---is introduced and used as a benchmark. Moreover, we investigate the robustness of the proposed tracking algorithms against outliers and examine their performance under different amounts of clutter.

Other distributed solutions for \ac{mtt} in a multi-sensor setup using the \ac{phd} filter have been studied, e.g.,  in \cite{Uney2013}, \cite{Uney2010}, \cite{battistelli2013consensus}. Contrary to our approach, they either assume overlapping \ac{fovs} or employ a pairwise communication scheme. The common idea, however, is to extend the single-sensor PHD filter to the multi-sensor case through communication between multiple nodes, or nodes and a fusion center. A more rigorous approach for MTT with multiple sensors is to use a multi-sensor PHD filter \cite{mahler2009the-multisensor, mahler2010approximate}, which seeks to estimate and track a single multi-sensor PHD instead of multiple single-sensor PHDs. In this work, we compare our methods to the approach in \cite{Uney2010} (adapted to our scenario), which is also based on single-sensor PHDs. The consideration of methods based on a multi-sensor PHD will be the focus of future work.

The paper is organized as follows: Section 2 presents the considered state-space model and recapitulates the theory of \acs{rfs}s as well as the \acs{phd} and the \acs{phd} filter. The problem of distributed \ac{mtt} is addressed in Section 3. Here, we will first detail our modification of \acl{atb} (\acs{atb}) before formulating the \acs{phd-dpf} and investigate its computational complexity and communication load. In Section 4, the \acs{ms-pphdf} is developed and analyzed in terms of computational complexity and communication load. Section 5 is dedicated to simulations. First, the \ac{dpcrlb} is introduced. Then, we present the simulation setup and discuss our results. Finally, a conclusion is given in Section 6.

\section{Models and Theory}

\subsection{State-Space and Measurement Model}
A linear state-space model is considered for each target at time instant $i \geq 0$. The target state vector $\boldsymbol{s}^\text{tgt}(i) = [\boldsymbol{x}^\text{tgt}(i), \boldsymbol{\dot{x}}^\text{tgt}(i)]^\top$ contains the target location vector $\boldsymbol{x}^\text{tgt}$ as well as the velocity vector $\boldsymbol{\dot{x}}^\text{tgt}$. For the sake of simplicity, we restrict ourselves to a 2D-environment. The target state evolves according to the state equation \cite{gustafsson2002particle}:
\begin{align}
\boldsymbol{s}^\text{tgt}(i) &= \boldsymbol{F}(i) \boldsymbol{s}^\text{tgt}(i-1) + \boldsymbol{G}(i) \boldsymbol{n}^\text{tgt}(i).
\label{eq:sysmod_lin}
\end{align}
The matrices $\boldsymbol{F}$ and $\boldsymbol{G}$ as well as the vector $\boldsymbol{n}^\text{tgt}$ will be explained shortly. Node $k$ obtains a measurement $\boldsymbol{z}_k$ of the target location as given by the measurement equation \cite{gustafsson2002particle}:
\begin{align}
\boldsymbol{z}_k(i) &= \boldsymbol{H}_k(i) \boldsymbol{s}^\text{tgt}(i) + \boldsymbol{\nu}_k^\text{tgt}(i),\qquad k \in \mathcal{M}
\label{eq:kf_leas}
\end{align}
with $\mathcal{M} = \left\{ m \in \left\{1, \ldots, N\right\}\ |\ \|\boldsymbol{x}_m(i) - \boldsymbol{x}^\text{tgt}(i)\|_2 \leq R_\text{sen} \right\}$ denoting the set of all nodes $m$ that are located such that the Euclidean distance $\|\boldsymbol{x}_m(i) - \boldsymbol{x}^\text{tgt}(i)\|_2$ between their location $\boldsymbol{x}_m$ and the target location $\boldsymbol{x}^\text{tgt}$ is not greater than their sensing radius $R_\text{sen}$. Note that $N$ is the total number of nodes in the network. Furthermore, $\boldsymbol{n}^\text{tgt}(i) \sim \mathcal{N}(\boldsymbol{0}_{2,1},\boldsymbol{Q}(i))$ and $\boldsymbol{\nu}_k^\text{tgt}(i) \sim \mathcal{N}(\boldsymbol{0}_{2,1},\boldsymbol{R}_k(i))$ denote the state and measurement noise processes, respectively, with the zero-mean vector $\boldsymbol{0}_{2,1} = \left[ 0, 0 \right]^\top$. Both noise processes are spatially and temporally white, as well as uncorrelated with the initial target state $\boldsymbol{s}^\text{tgt}(0)$ and each other for all $i$. For the sake of simplicity, we choose a time-invariant measurement noise covariance matrix
\begin{align}
\boldsymbol{R}_k(i) = \boldsymbol{R}_k = \sigma_r^2 \boldsymbol{I}_2,
\end{align}
where $\sigma_r^2$ is the variance of each component of the measurement noise and $\boldsymbol{I}_n$ denotes the identity matrix of size $n$.

In target tracking, the model matrices are usually chosen to be time-invariant and given by \cite{gustafsson2002particle}
\begin{equation}
\boldsymbol{F} = \begin{bmatrix} \boldsymbol{I}_2 & \Delta i \boldsymbol{I}_2\\ \boldsymbol{0}_{2,2} & \boldsymbol{I}_2\end{bmatrix},\qquad \boldsymbol{G} = \begin{bmatrix}\frac{\Delta i^2}{2} \boldsymbol{I}_2 \\ \Delta i \boldsymbol{I}_2 \end{bmatrix},\qquad \boldsymbol{Q} = \sigma_q^2 \boldsymbol{I}_2,
\end{equation}
where $\boldsymbol{0}_{2,2}$ is the $2 \times 2$ zero matrix. Furthermore, $\Delta i$ is the time step interval in seconds with which the state-space model progresses. In addition, $\sigma_q^2$ denotes the variance of a state noise component. We assume that the sensor nodes only obtain information on the location of a target. One common set of measurements that is often found in applications at sea is the combination of distance and bearing measurements from which an estimate of the target location can easily be calculated. Since we are not interested in the exact nature of the measured location information but rather in how this information is processed by different tracking algorithms, we formulate our measurement model based on the local target location estimates at each node. This gives us a general model that is applicable to a wide variety of application irrespective of the exact measurement quantities. Thus, we obtain a general measurement matrix $\boldsymbol{H}_k$ of the form
\begin{equation}
\boldsymbol{H}_k = \begin{bmatrix} \boldsymbol{I}_2 & \boldsymbol{0}_{2,2} \end{bmatrix}.
\end{equation}

\subsection{Random Finite Sets (RFSs)}
A \ac{rfs} is an unordered finite set that is random in the number of its elements as well as in their values \cite{vo2008random,mahler2007statistical,mahler2014advances}. Therefore, \ac{rfs}s are a natural choice for representing the multi-target states and measurements in \ac{mtt}: the state and measurement vectors of all targets are collected in corresponding \ac{rfs}s \cite{vo2003sequential, vo2005sequential}. Given the realization $\Xi_{i-1}$ of the \ac{rfs} $\boldsymbol{\Xi}_{i-1}$ at time instant $i-1$, the multi-target state of our tracking problem can be described by the \ac{rfs} $\boldsymbol{\Xi}_i$ according to
\begin{align}
\boldsymbol{\Xi}_i = \boldsymbol{\mathcal{S}}_i(\Xi_{i-1}) \cup \boldsymbol{\mathcal{B}}_i,
\end{align}
where the survival set $\boldsymbol{\mathcal{S}}_i(\Xi_{i-1})$ denotes the \ac{rfs} of targets that already existed at time step $i-1$ and have not exited the \ac{roi}, i.e., the region covered by the sensor network, in the transition to time step $i$. In addition, the birth set $\boldsymbol{\mathcal{B}}_i$ is the \ac{rfs} of new targets that spontaneously appear at the border of the \ac{roi} at time instant $i$ \cite{challa2011fundamentals, mahler2003multitarget, vo2005sequential}. Note that the statistical behavior of $\boldsymbol{\Xi}_i$ can be described by the conditional probability $f_{i|i-1}(\Xi_i | \Xi_{i-1})$. 

The multi-target measurement model is given by the \ac{rfs} $\boldsymbol{\Sigma}_i$ as
\begin{align}
\boldsymbol{\Sigma}_i = \boldsymbol{\Theta}_i(\Xi_i) \cup \boldsymbol{\mathcal{C}}_i(\Xi_i),
\end{align}
where $\boldsymbol{\Theta}_i(\Xi_i)$ is the \ac{rfs} of measurements generated by $\Xi_i$. In addition, the \ac{rfs} $\boldsymbol{\mathcal{C}}_i(\Xi_i)$ represents clutter or false alarms. Given a realization $\Sigma_i$ of $\boldsymbol{\Sigma}_i$, the statistical behavior of  the \ac{rfs} $\boldsymbol{\Sigma}_i$ is described by the conditional probability $f_i(\Sigma_i | \Xi_i)$.

\subsection{The Probability Hypothesis Density (PHD)}
In analogy to the single-target case, the optimal Bayesian filter for \ac{mtt} recursively propagates the multi-target posterior $f_{i|i}(\Xi_i | \Sigma_{0:i})$ over time, according to
\begin{align}
f_{i|i}(\Xi_i | \Sigma_{0:i}) &= \frac{f_i(\Sigma_i | \Xi_i) f_{i|i-1}(\Xi_i | \Sigma_{0:i-1})}{\int f_i(\Sigma_i | \Xi) f_{i|i-1}(\Xi | \Sigma_{0:i-1}) \mu_s(d\Xi)}\\
f_{i|i-1}(\Xi_i | \Sigma_{0:i-1}) &= \int f_{i|i-1}(\Xi_i | \Xi) f_{i-1|i-1}(\Xi | \Sigma_{0:i-1}) \mu_s(d\Xi),
\end{align}
where $\mu_s$ is a dominating measure as described in \cite{vo2005sequential}. This approach requires the evaluation of multiple integrals, which makes it even more computationally challenging than its single-target counterpart. A common solution is to find a set of statistics, e.g., the moments of first or second order, which yield a good approximation of the posterior, and propagate them instead \cite{challa2011fundamentals}.

The \acl{phd} (\acs{phd}) $D_{i|i}(\boldsymbol{s}(i)|\Sigma_{0:i})$ is an indirect first-order moment of $f_{i|i}(\Xi_i | \Sigma_{0:i})$ \cite{mahler2001multitarget}. It is given by the following integral \cite{mahler2003multitarget,mahler2013statistics}:
\begin{align}
\begin{aligned}
D_{i|i}(\boldsymbol{s}(i)|\Sigma_{0:i}) &= \sum_{\boldsymbol{s}_n^\text{tgt}(i) \in \Xi_i} \int\delta(\boldsymbol{s}(i) - \boldsymbol{s}_n^\text{tgt}(i)) f_{i|i}(\boldsymbol{s}(i) | \Sigma_{0:i})  d\boldsymbol{s}(i),
\end{aligned}
\end{align}
where $\int f(Y)\delta Y$ denotes a set integral. 

The \ac{phd} has the following two properties \cite{mahler2001multitarget}:

\begin{enumerate}
\item The expected number of targets $\hat{N}_\text{tgt}(i)$ at time step $i$ is obtained by integrating the \ac{phd} according to
\begin{align}
\hat{N}_\text{tgt}(i) = \int D_{i|i}(\boldsymbol{s}(i)|\Sigma_{0:i}) d\boldsymbol{s}(i).
\end{align}
This is in contrast to \acl{pdf}s (\acs{pdf}s), which always integrate to 1.
\item Estimates of the individual target states can be found by searching for the $\left\lfloor \hat{N}_\text{tgt} \right\rceil$ highest peaks of the PHD, where $\left\lfloor\cdot\right\rceil$ denotes rounding to the nearest integer.
\end{enumerate}
Because of these two properties, the number of targets as well as their states can be estimated independently at each time step without any knowledge of their identities. That way, the data association issue is avoided. However, this also means that \ac{phd} Filters cannot deliver the continuous track of a specific target. If continuous tracks are required, an additional association step has to be performed. Two possible association algorithms for track continuity can be found in \cite{clark2006multiple}.

%

\subsection{The PHD Filter}
The \ac{phd} Filter is an approach for recursively propagating the \ac{phd} $D_{i|i}(\boldsymbol{s}(i)|\Sigma_{0:i})$ at time step $i$ given measurements up to time step $i$ over time. If the \ac{rfs} $\boldsymbol{\Xi}$ is Poisson-distributed, then its \ac{phd} is equal to its intensity function and is, hence, a sufficient statistic \cite{mahler2003multitarget}. In this case, the PHD recursion is given by the following prediction and update equations \cite{mahler2003multitarget}:
\begin{align}
D_{i|i-1}(\boldsymbol{s}(i)|\Sigma_{0:i-1}) &= b_i(\boldsymbol{s}(i)) + \int p_S(\boldsymbol{s}(i-1)) f_{i|i-1}(\boldsymbol{s}(i)|\boldsymbol{s}(i-1)) D_{i-1|i-1}(\boldsymbol{s}(i-1)|\Sigma_{0:i-1}) d\boldsymbol{s}(i-1)\label{eq:phd_prediction}\\
D_{i|i}(\boldsymbol{s}(i)|\Sigma_{0:i}) &= \Bigg[1 - p_D + \sum_{\boldsymbol{z} \in \Sigma_i} \frac{p_D f_i(\boldsymbol{z}|\boldsymbol{s}(i))}{\lambda_\text{FA} c_\text{FA}(\boldsymbol{z}) + p_D \int f_i(\boldsymbol{z}|\boldsymbol{s}(i)) D_{i|i-1}(\boldsymbol{s}(i)|\Sigma_{0:i-1}) d\boldsymbol{s}(i)} \Bigg] D_{i|i-1}(\boldsymbol{s}(i)|\Sigma_{0:i-1})\label{eq:phd_update}
\end{align}
Note that $b_i(\boldsymbol{s}(i))$ is the \ac{phd} of the birth set $\boldsymbol{\mathcal{B}}_i$ of new targets appearing at time step $i$. In addition, $p_S(\boldsymbol{s}(i-1))$ denotes the probability that a target survives the transition from time step $i-1$ to $i$. The probability of survival depends on the previous state $\boldsymbol{s}(i-1)$ because a target that is close to the border of the \ac{roi} and has a velocity vector pointing away from it is unlikely to be present at time step $i$. Furthermore, $f_{i|i-1}(\boldsymbol{s}(i)|\boldsymbol{s}(i-1))$ and $f_i(\boldsymbol{z}|\boldsymbol{s}(i))$ denote the transition probability and the likelihood, respectively. The probability of detection $p_D$ is constant over time and the tracker's \ac{fov} since it is assumed that all targets can be detected if the \ac{roi} is covered. The term $\lambda_\text{FA} c_\text{FA}(\boldsymbol{z})$ represents Poisson-distributed false-alarms due to clutter, where $\lambda_\text{FA}$ is the false alarm parameter, which is distributed according to its spatial distribution $c_\text{FA}(\boldsymbol{z})$. 

\section{Distributed Multi-Target Target Tracking}

In this section we introduce the \acl{phd-dpf} (\acs{phd-dpf}), a distributed Particle Filter implementation of the PHD Filter for performing \ac{mtt} in a sensor network without a fusion center. Before diving into the algorithm, we briefly review the concept of \ac{atb} and discuss the modification we applied in the \acs{phd-dpf}.
 
\subsection{Adaptive Target Birth (ATB)}
Standard formulations of the PHD Filter consider the PHD $b_i(\boldsymbol{s}(i))$ of the birth set $\boldsymbol{\mathcal{B}}_i$ to be known a priori \cite{Ristic2012}. For typical tracking applications such as air surveillance, this is a reasonable assumption since new targets should appear at the border of the \ac{roi} given continuous observation. An alternative is to make the target birth process adaptive and measurement-driven as suggested in \cite{Ristic2012, Ristic2010}. To this end, the PHD---and consequently the set of particles and weights approximating it in a Particle Filter implementation---is split into two densities corresponding to persistent objects, which have survived the transition from time step $i-1$ to $i$, and newborn objects, respectively.

In \cite{Ristic2012, Ristic2010}, the PHD of newborn objects is approximated by randomly placing $N_P$ new particles around each target measurement, with $N_P$ denoting the number of particles per target. We improve upon this approach by only considering measurements with no noticeable impact on any persistent particle weight, as these may indicate the appearance of a new target. That way, the number of newborn particles is further reduced and a possible overlap between persistent and newborn PHD is avoided. With the transition to time step $i+1$, the newborn particles become persistent. Furtherore, we perform the ATB step towards the end of each iteration of the algorithm and only consider the particles representing the persistent PHD in the prediction, weighting, and resampling steps. Hence, the update equation \eqref{eq:phd_update} does not have to be modified as in \cite{Ristic2012, Ristic2010}.

While \ac{atb} delays the tracking algorithm by one time step, it is much more efficient as it only places new particles in regions in which a target is likely to be found. In addition, there is no need for an explicit  initialization step since the first incoming target will trigger the deployment of a newborn particle cloud around its corresponding measurement.

\subsection{The Diffusion Particle PHD Filter (D-PPHDF)}
The proposed \acl{phd-dpf} (\acs{phd-dpf}) is an extension of the single-sensor \ac{phd-pf} \cite{clark2006multiple, vo2003sequential, hong2011simplified} for the multi-sensor case. Furthermore, it relies on \ac{atb} for a more efficient target detection. The communication scheme we employ to exchange measurements and estimates between nodes is inspired by the two-step communication used in the context of Diffusion Adaptation \cite{sayed2012diffusion}. However, the algorithm does not rely on least-mean-squares or any other kind of adaptive filter. First, each node $k$ in the active part of the network obtains an intermediate estimate of the states of the targets present, i.e., of the PHD of persistent targets---represented by the set $\left\{\boldsymbol{s}^p_{k,\text{pers}}(i),w^p_{k,\text{pers}}(i)\right\}_{p=1}^{N_{k,\text{pers}}(i)}$ of persistent particles with corresponding weights---based on neighborhood measurements. In other words, every active node runs a separate \ac{phd-pf} with access to measurements from its neighborhood $\mathcal{N}_k$, defined as
\begin{align}
\mathcal{N}_k =\{l \in \{1,\ldots,N\}\;\big|\;\|\boldsymbol{x}_l-\boldsymbol{x}_k\|_2 \leq R_\text{com}\}, \;\;k= 1,\ldots,N.,
\label{eqn:neighborhood}
\end{align}
where $R_\text{com}$ denotes the communication radius. Second, each active node combines the intermediate estimates from its neighborhood to a final, collaborative estimate. To this end, the persistent particle sets of all neighbors are merged into a collective set $\left\{\boldsymbol{s}^p_{k,\text{coll}}(i),w^p_{k,\text{coll}}(i)\right\}_{p=1}^{N_{k,\text{coll}}(i)}$ of persistent neighborhood particles and corresponding weights before the clustering step, with $N_{k,\text{coll}}(i)$ denoting the number of collective persistent neighborhood particles. In the sequel, we will look at the individual steps of the \ac{phd-dpf} in more detail:

\begin{itemize}
\item \emph{Merging}: The sets $\left\{\boldsymbol{s}^p_{k,\text{coll}}(i-1),w^p_{k,\text{coll}}(i-1)\right\}_{p=1}^{N_{k,\text{coll}}(i-1)}$ and $\left\{\boldsymbol{s}^p_{k,\text{new}}(i-1),w^p_{k,\text{new}}(i-1)\right\}_{p=1}^{N_{k,\text{new}}(i-1)}$ consist of the collective persistent neighborhood particles and newborn particles of node $k$, $\boldsymbol{s}^p_{k,\text{coll}}(i-1)$ and $\boldsymbol{s}^p_{k,\text{new}}(i-1)$, respectively, at time step $i-1$ with their respective weights $w^p_{k,\text{coll}}(i-1)$ and $w^p_{k,\text{new}}(i-1)$. These sets are merged to become the total set $\left\{\boldsymbol{s}^p_{k,\text{tot}}(i),w^p_{k,\text{tot}}(i)\right\}_{p=1}^{N_{k,\text{tot}}(i)}$ of particles and weights of node $k$ at time step $i$. Here, $N_{k,\text{tot}}(i)$ is the total number of particles of node $k$ at time step $i$, which is given by
\begin{align}
N_{k,\text{tot}}(i) = N_{k,\text{coll}}(i-1) + N_{k,\text{new}}(i-1),
\end{align}
with $N_{k,\text{coll}}(i-1)$ and $N_{k,\text{new}}(i-1)$ denoting the respective number of persistent neighborhood and newborn particles at the previous time step. Note that since the sets of particles and weights represent PHDs, merging the sets corresponds to summing these PHDs.

\item \emph{Predicting:}
Each particle is propagated through the system model to become a persistent particle. The system model is assumed to be the same for each target and given by Equation~(\ref{eq:sysmod_lin}). Since the process noise is captured by the spread of the particle cloud, the respective term can be removed from the equation, yielding
\begin{align}
\boldsymbol{s}^p_{k,\text{pers}}(i) = \boldsymbol{F} \boldsymbol{s}^p_{k,\text{tot}}(i), \qquad p = 1, \ldots, N_{k,\text{tot}}(i).
\label{eq:d_pred1}
\end{align}
The corresponding weights are multiplied with the probability of survival $p_S$, which is assumed to be constant for the sake of simplicity\footnote{A constant probability of survival $p_S$ is a reasonable assumption if the targets move relatively slowly with respect to the observation time and the size of the \ac{roi}.}, according to
\begin{align}
\begin{aligned}
w^p_{k,\text{pers}}(i|i-1) &= p_S w^p_{k,\text{tot}}(i), \qquad p = 1, \ldots, N_\text{pers}(i).
\end{aligned}
\label{eq:d_pred2}
\end{align}

The prediction of particles and weights corresponds to the second term in Equation~(\ref{eq:phd_prediction}). 

%

\item \emph{Measuring \& Broadcasting (1):} The sensor nodes obtain measurements of the targets and forward them to their neighbors.


\item \emph{Weighting:}
The persistent particle weights of node $k$ are updated by applying a weighting step corresponding to Equation (\ref{eq:phd_update}) iteratively for each neighbor. Using the product operator, this weighting step can be compactly denoted as

\begin{align}
\begin{aligned}
w_{k,\text{pers}}^p(i) &= \prod_{l \in \mathcal{N}_k}\Bigg[1 - p_D + \sum_{\boldsymbol{z}_j \in \Sigma_i^l} w^p_{k,j,\text{update}}(i) \Bigg] w_{k,\text{pers}}^p(i|i-1),\label{eq:dphdpf_update}
\end{aligned}
\end{align}
with
\begin{align}
w^p_{k,j,\text{update}}(i) &= \frac{p_D f_i(\boldsymbol{z}_j\;|\;\boldsymbol{x}^p(i))}{\lambda_\text{FA} c_\text{FA}(\boldsymbol{z}_j) +  \mathcal{L}(\boldsymbol{z}_j)},
\label{eq:d_weight2}
\end{align}

where $\Sigma_i^l$ is the set of measurements obtained by node $l$ and $\mathcal{L}(\boldsymbol{z}_j)$ is calculated as 
\begin{align}
\mathcal{L}(\boldsymbol{z}_j) = \sum_{q = 1}^{N^p_{k,\text{pers}}(i)} p_D f_i(\boldsymbol{z}_j\;|\;\boldsymbol{x}^q(i)) w^q_{k,\text{pers}}(i|i-1).\label{eq:d_weight3}
\end{align}
Note that $f_i(\boldsymbol{z}_j\;|\;\boldsymbol{x}^p(i))$ is the likelihood and $\boldsymbol{x}^p(i)$ is the location vector of particle $p$.

Afterwards, each node $k$ obtains the set $\Sigma_{i,\text{cand}}^k$ of candidate measurements, i.e., measurements that are not responsible for the highest weighting of any persistent particle, to be used in the \acs{atb} step later on. The set $\Sigma_{i,\text{cand}}^k$ is found according to
\begin{align}
\begin{aligned}
	\Sigma_{i,\text{cand}}^k = \Sigma_i^k \backslash \Big\{\boldsymbol{z}_{m_p}\;\Big|\;m_p &= \arg \underset{j}{\max}\ w^p_{k,j,\text{update}}(i), p = 1,...,N_{k,\text{pers}}(i)\Big\}.
	\end{aligned}
	\label{eq:dpphdf_cand}
\end{align}

\item \emph{Resampling:} Each node $k$ calculates its own expected number of targets $\hat{N}_{k,\text{tgt}}(i)$ from its total persistent particle mass according to
\begin{align}
\hat{N}_{k,\text{tgt}}(i) = \left\lfloor \sum_{p = 1}^{N_{k,\text{tot}}(i)} w^p_{k,\text{pers}}(i) \right\rceil.
\label{eq:dpphdf_ntgt}
\end{align}
Consequently, the number of persistent particles of node $k$ is updated as
\begin{align}
N_{k,\text{pers}}(i) = \hat{N}_{k,\text{tgt}}(i) N_P.
\end{align}
Furthermore, the set of persistent particles of node $k$ has to be resampled by drawing $N_{k\text{pers}}(i)$ particles with replacement from it. Note that the probability of drawing particle $p$ is given by $\frac{w^p_{k,\text{pers}}(i)}{\hat{N}_{k,\text{tgt}}(i)}$ since the weights do not sum to unity. Then, the weights are reset to equal values as
\begin{align}
w^p_{k,\text{pers}}(i) = \frac{\hat{N}_{k,\text{tgt}}(i)}{N_{k,\text{pers}}(i)}, \qquad p = 1, \ldots, N_{k,\text{pers}}(i).
\label{eq:dpphdf_weight}
\end{align}

\item \emph{Broadcasting (2):} Every node $k$ transmits its set of resampled persistent particles and weights $\left\{\boldsymbol{s}^p_{k,\text{pers}}(i),w^p_{k,\text{pers}}(i)\right\}_{p=1}^{N_{k,\text{pers}}(i)}$ to its neighbors.

\item \emph{Clustering:} Each node $k$ forms a collective set of persistent neighborhood particles $\boldsymbol{s}^p_{k,\text{coll}}(i)$ and corresponding weights $w^p_{k,\text{Nh}}(i)$ according to
\begin{align}
\begin{aligned}
&\left\{\boldsymbol{s}^p_{k,\text{coll}}(i),w^p_{k,\text{coll}}(i)\right\}_{p=1}^{N_{k,\text{coll}}(i)} = \bigcup_{l \in \mathcal{N}_k} \left\{\boldsymbol{s}^p_{l,\text{pers}}(i),w^p_{l,\text{pers}}(i)\right\}_{p=1}^{N_{l,\text{pers}}(i)},
\end{aligned}
\end{align}
with
\begin{align}
N_{k,\text{coll}}(i) = \sum_{l \in \mathcal{N}_k} N_{l,\text{pers}}(i)
\end{align}
denoting the number of collective persistent neighborhood particles of node $k$. As in the \emph{merging} step, this corresponds to summing the corresponding PHDs to obtain an updated single-sensor PHD with a probability distribution reflecting the information of the entire neighborhood of node $k$. Note that the PHDs might not be independent if a target is detected by more than one neighbor. However, this is not a problem since merging the particle sets simply results in the respective target being represented by more particles. Hence, node $k$ will be able to estimate the corresponding location more accurately.

The estimated target states are found by clustering the collective persistent particles. Since the expected number of targets $\hat{N}_{l,\text{tgt}}(i),\ l \in \mathcal{N}_k$ might be different for each neighbor, we resort to \emph{hierarchical clustering of the single-linkage type} \cite{everitt2011cluster}. Here, the sum of the expected number of targets over the neighborhood can serve as an upper bound for the number of clusters. Note, however, that if two targets are close to each other, clustering algorithms might not be able to resolve both targets correctly.

\item \emph{Roughening:} A roughening step is performed to counter sample impoverishment \cite{gordon1993novel}. To this end, an independent jitter $\boldsymbol{s}^j(i)$ is added to every resampled particle. Each component $s_c^j(i),\;c = 1, \ldots, d$ of the jitter with dimensionality $d$ is sampled from the Gaussian distribution $\mathcal{N}(0,(\sigma_c^j(i))^2)$. The component-wise standard deviation of the jitter is given by
\begin{align}
\sigma_c^j(i) = K E_c N_{k,\text{coll}}(i)^{-1/d},
\label{eq:roughening}
\end{align}
where $E_c$ is the interval length between the maximum and minimum samples of the respective component. To avoid evaluating $E_c$ separately for each particle cluster, it is assigned an empirically found constant value.\footnote{Since the noise variances as well as the network topology are fixed, the true value of $E_c$ will not change significantly over time and between clusters, so this is a valid simplification.} Note that $d = 4$ since the dimensionality of the jitter vector $\boldsymbol{s}^j(i)$ and the particle state vector $\boldsymbol{s}^p(i)$ have to coincide. In addition, $K$ is a tuning constant, which controls the spread of the particle cloud.

\item \emph{Adaptive Target Birth:} $N_P$ new particles are placed randomly around each candidate measurement $\boldsymbol{z}_j \in \Sigma_{i,\text{cand}}^k$ leading to a total number of $N_{k,\text{new}}(i) = N_P\cdot|\Sigma_{i,\text{cand}}^k|$ newborn particles for node $k$. Every newborn particle is associated with a weight that is chosen according to
\begin{align}
w^p_{k,\text{new}}(i) = \frac{p_B}{N^p_{k,\text{new}}(i)}, \qquad p = 1, \ldots, N_{k,\text{new}}(i),
\label{eq:dpphdf_atb}
\end{align}
where $p_B$ is the probability of birth. Depending on the application, $p_B$ can depend on time as well as on the location of the respective particle. For simplicity, the probability that a new target enters the \ac{roi} is assumed to be equal for all locations in the birth region over time. The target birth process corresponds to the first term in Equation~(\ref{eq:phd_prediction}).
\end{itemize}

Figure \ref{fig:subtracks}) shows an example of tracking three targets, which move along the deterministic tracks depicted in Figure~\ref{fig:mtt_tracks}), using the \ac{phd-dpf}. Note that each small colored dot corresponds to a target location estimate obtained by the respective node with the same color while the light grey dots represent the collective measurements from all nodes. From this illustration, the following properties of the \ac{phd-dpf} are apparent: First, the algorithm only delivers separate location estimates~--~represented by the small colored dots~--~for each time instant rather than continuous tracks, which~--~as mentioned before~--~is a common property of \ac{phd} filters. Second, the network as a whole would be able to correctly track all three targets, while a single node only obtains the locally relevant subtracks of the targets in its vicinity. Third, the employed two-step communication scheme is able to extend the vicinity of a node far beyond its own sensing radius of $R_\text{sen} = 6~\textrm{m}$. This can, for instance, be seen from the fact that the lime-green node located at $[-14, -23]$ is able to obtain location estimates of target 2, which enters the \ac{roi} from the south. Finally, Figure~\ref{fig:subtracks}) also illustrates the resolution problem of clustering. When targets 1 and 2, which enter the \ac{roi} from the north and the south, respectively, cross paths, the nodes in their vicinity see them as just one target. This leads to an aggregation of target location estimates around $[9, 0]$.

	The pseudo-code of the \ac{phd-dpf} is given in Table \ref{alg:d-pphdf}.
	
	\begin{table*}[p]
  \centering
  \begin{framed}
    \begin{algorithmic}[1]
      \STATE \textbf{input:} $d, E_c, K, n, N, N_P, p_B, p_S, \lambda_\text{FA}, c_\text{FA}$ 
      \STATE \textbf{initialize:} $\left\{\boldsymbol{s}^p_{k,\text{coll}}(0),w^p_{k,\text{coll}}(0)\right\}_{p=1}^{N_{k,\text{coll}}(0)} =\left\{\boldsymbol{s}^p_{k,\text{new}}(0),w^p_{k,\text{new}}(0)\right\}_{p=1}^{N_{k,\text{new}}(0)} = \emptyset $.
      \WHILE{$i\le n$}
      \FOR{$k=1,\ldots,N$}
        \STATE Merge the sets of collective persistent and newborn particles with corresponding weights:
        \begin{equation*}
        	\left\{\boldsymbol{s}^p_{k,\text{tot}}(i),w^p_{k,\text{tot}}(i)\right\}_{p=1}^{N_{k,\text{tot}}(i)} = \left\{\boldsymbol{s}^p_{k,\text{coll}}(i-1),w^p_{k,\text{coll}}(i-1)\right\}_{p=1}^{N_{k,\text{coll}}(i-1)} \cup \left\{\boldsymbol{s}^p_{k,\text{new}}(i-1),w^p_{k,\text{new}}(i-1)\right\}_{p=1}^{N_{k,\text{new}}(i-1)}.
        \end{equation*}
		\FOR{$p=1,\ldots,N_{k,\text{tot}}(i)$}
			\STATE Predict the new state of each particle and update the weight with the probability of survival $p_S$:
			\begin{align*}
\boldsymbol{s}^p_{k,\text{pers}}(i) &= \boldsymbol{F} \boldsymbol{s}^p_{k,\text{tot}}(i)\\
w^p_{k,\text{pers}}(i|i-1) &= p_S w^p_{k,\text{tot}}(i).
\end{align*}
			\STATE Update the weights using neighborhood measurements:

\begin{align*}
w_{k,\text{pers}}^p(i) &= \prod_{l \in \mathcal{N}_k}\Bigg[1 - p_D +\sum_{\boldsymbol{z}_j \in \Sigma_i^l} w^p_{k,j,\text{update}}(i) \Bigg] w_{k,\text{pers}}^p(i|i-1),\\
w^p_{k,j,\text{update}}(i) &= \frac{p_D f_i(\boldsymbol{z}_j\;|\;\boldsymbol{x}^p(i))}{\lambda_\text{FA} c_\text{FA}(\boldsymbol{z}_j) +  \mathcal{L}(\boldsymbol{z}_j)},\\
\mathcal{L}(\boldsymbol{z}_j) &= \sum_{q = 1}^{N_{k,\text{pers}}(i)} p_D f_i(\boldsymbol{z}_j\;|\;\boldsymbol{x}^q(i)) w^q_{k,\text{pers}}(i|i-1).
\end{align*}
		\ENDFOR
		\STATE Form the set of candidate measurements for ATB:
		\begin{align*}
	\Sigma_{i,\text{cand}}^k = \Sigma_i^k \backslash \Big\{\boldsymbol{z}_{m_p}\;\Big|\;m_p &= \arg \underset{j}{\max}\ w^p_{k,j,\text{update}}(i),\quad p = 1,...,N_{k,\text{pers}}(i), \quad j = 1,...,\left|\Sigma_i^l\right| \forall l\in\mathcal{N}_k\Big\}.
\end{align*}
		\STATE Calculate the estimated number of targets:
		\begin{align*}
\hat{N}_{k,\text{tgt}}(i) = \left\lfloor \sum_{p = 1}^{N_{k,\text{tot}}(i)} w^p_{k,\text{pers}}(i) \right\rceil.
\end{align*}
        \STATE Resample $\hat{N}_{k,\text{pers}}(i) = \hat{N}_{k,\text{tgt}}(i)N_P$ particles and reset the weights:
        \begin{align*}
w^p_{k,\text{pers}}(i) = \frac{\hat{N}_{k,\text{tgt}}(i)}{N_{k,\text{pers}}(i)}, \qquad p = 1, \ldots, N_{k,\text{pers}}(i).
\end{align*}
		\STATE Merge the sets of persistent neighborhood particles and weights:
        \begin{equation*}
        	\left\{\boldsymbol{s}^p_{k,\text{coll}}(i),w^p_{k,\text{coll}}(i)\right\}_{p=1}^{N_{k,\text{coll}}(i)} = \bigcup_{l \in \mathcal{N}_k} \left\{\boldsymbol{s}^p_{l,\text{pers}}(i),w^p_{l,\text{pers}}(i)\right\}_{p=1}^{N_{l,\text{pers}}(i)}.
        \end{equation*}
        \STATE Use single-linkage clustering to identify $\hat{N}_{\text{tgt}}(i)$ clusters and find the set of estimated target states $\left\{\hat{\boldsymbol{s}}^l_k(i)\right\}_{l=1}^{\hat{N}_{\text{tgt}}(i)}$ by calculating the centroids.
        \STATE Add an independent jitter to each particle using a component-wise standard deviation of:
        \begin{align*}
\sigma_c^j(i) = K E_c N_{k,\text{coll}}(i)^{-1/d}.
\end{align*}

        \STATE Place $N_P$ new particles randomly around each candidate measurement $\boldsymbol{z}_j \in \Sigma_{i,\text{cand}}^k$. Set the weights as:
        \begin{align*}
w^p_{k,\text{new}}(i) = \frac{p_B}{N_{k,\text{new}}(i)}, \qquad p = 1, \ldots, N_{k,\text{new}}(i).
\end{align*}
      \ENDFOR
      \STATE $i\gets i+1$ 
      \ENDWHILE
      \RETURN 
    \end{algorithmic}
  \end{framed}
  
  \caption{The Diffusion Particle PHD Filter.}
  \label{alg:d-pphdf} 
\end{table*}

\subsection{Computational Complexity and Communication Load}
In this section we take a look at the computational complexity and the communication load the \ac{phd-dpf} imposes on each node in the active subnetwork. The following steps are performed at every time instant $i$ but time dependency is omitted for simplicity. Note that each of the steps scales with the number of active nodes when considering the computational complexity of the network as a whole.
\begin{itemize}
	\item \emph{Prediction}: The prediction step described by Equations \eqref{eq:d_pred1} and \eqref{eq:d_pred2} is performed for each particle at every active node. Hence, it scales with the number of particles $N_{k,\text{tot}}$ and the dimensionality $d$ of the particle vectors. In order to obtain a tractable expression for the computational complexity, we assume each node to have the same number of particles $N_\text{tot}$.\\
	$\Rightarrow \mathcal{O}(N_\text{tot}d)$
	\item \emph{Weighting}: Each particle is updated in the weighting step given by Equations \eqref{eq:dphdpf_update}-\eqref{eq:dpphdf_cand}. The weight update as well as the designation of candidate measurements for ATB depends on the neighborhood size $\left|\mathcal{N}_k\right|$ of node $k$ and the number of measurements $\left|\Sigma^l\right|$ of each of its neighbors $l$. For tractability reasons, we assume each node to have the same number of neighbors $N_\text{nb}$ and to obtain the same number of measurements $N_\text{meas}$.\\
	$\Rightarrow \mathcal{O}(N_\text{tot}N_\text{nb}N_\text{meas})$ 
	\item \emph{Resampling}: The estimation of the number of targets and the resampling step in Equations \eqref{eq:dpphdf_ntgt}-\eqref{eq:dpphdf_weight} are linear in the number of particles used for the calculation \cite{gustafsson2010particle}. For the sake of simplicity, we assume each active node to have the same estimate of the number of targets $N_\text{tgt}$.\\
	$\Rightarrow \mathcal{O}(N_\text{tot} + N_\text{active}N_\text{tgt}N_P)$
	\item \emph{Clustering}: The complexity of single-linkage clustering is cubic in the number of particles, i.e., in the number of neighbors $N_\text{nb}$ of each node, the estimated number of targets $N_\text{tgt}$, the number of particles per target $N_P$, and the dimensionality $d$ of the particles \cite{murtagh1983survey}.\\ 
	$\Rightarrow \mathcal{O}((N_\text{nb}N_\text{tgt}N_Pd)^3)$
	\item \emph{Roughening}: Roughening (Equation \eqref{eq:roughening}) is performed for every collective particle and is linear in the dimensionality of the particles.\\
	$\Rightarrow \mathcal{O}(N_\text{nb}N_\text{tgt}N_Pd)$
	\item \emph{Adaptive Target Birth}: The birth process depends on the number of particles per target $N_P$ as well as the number of candidate measurements $N_\text{cand}$, which is assumed equal for each active node to ensure tractibility.\\
	$\Rightarrow \mathcal{O}(N_PN_\text{cand})$
	
	
\end{itemize}
As far as the communication load is concerned, the \ac{phd-dpf} requires the broadcasting of measurements, i.e., 2 scalars per measurement, over the neighborhood in the first broadcasting step. In the second step, the sets of particles and weights, i.e., 5 scalars per particle, are transmitted. Clearly, the communication load strongly depends on the number of nodes in the network, or more precisely the number of active nodes and the size of their respective neighborhood.  
As an extension of the \ac{phd-dpf}, one could think of changing the second broadcasting step and transmit Gaussian Mixture Model representations---instead of the actual particles and weights---that will be resampled at the receiver node (see e.g., \cite{vo2006gaussian}). That way, communication load could be reduced to transmitting only a few scalars in the second broadcasting step at the cost of estimation accuracy and additional computational complexity. However, a thorough treatment of this extension is beyond the scope of this work.

\section{Centralized Multi-Target Tracking}

Having presented the \ac{phd-dpf} as a distributed solution for \ac{mtt} in a sensor network, we propose the centralized counterpart to our approach in the sequel.

\subsection{The Multi-Sensor Particle PHD Filter (MS-PPHDF)}
The proposed \ac{ms-pphdf} is a centralized, multi-sensor \ac{phd-pf} that relies on a fusion center with access to the measurements of all nodes in the network. It is based on the formulation of the single-sensor \ac{phd-pf} in \cite{clark2006multiple, vo2003sequential, hong2011simplified} but with an extended measurement set comprising the measurements of the entire network. Hence, one might obtain more than one measurement per target---a change to the typical assumption in target tracking that each target produces \emph{at most} one measurement \cite{bar-shalom2011tracking}. To account for this change, we add a pre-clustering step before the weighting step and normalize the weight update accordingly. A similar partitioning of the measurement set is used in extended target tracking, where a sensor can receive multiple target reflections due to the target's physical extent \cite{granstrom2012, granstrom2014}.

	In the following, we will look at the individual steps of the algorithm in more detail:

\begin{itemize}
\item \emph{Merging}: The sets $\left\{\boldsymbol{s}^p_\text{pers}(i-1),w^p_\text{pers}(i-1)\right\}_{p=1}^{N_\text{pers}(i-1)}$ and $\left\{\boldsymbol{s}^p_\text{new}(i-1),w^p_\text{new}(i-1)\right\}_{p=1}^{N_\text{new}(i-1)}$ consist of the persistent and newborn particles, $\boldsymbol{s}^p_\text{pers}(i-1)$ and $\boldsymbol{s}^p_\text{new}(i-1)$, respectively, at time step $i-1$ with their respective weights $w^p_\text{pers}(i-1)$ and $w^p_\text{new}(i-1)$. These sets are merged to become the total set $\left\{\boldsymbol{s}^p_\text{tot}(i),w^p_\text{tot}(i)\right\}_{p=1}^{N_\text{tot}(i)}$ of particles and weights at time step $i$. Here, $N_\text{tot}(i)$ is the total number of particles at time step $i$, which is given by
\begin{align}
N_\text{tot}(i) = N_\text{pers}(i-1) + N_\text{new}(i-1),
\end{align}
with $N_\text{pers}(i-1)$ and $N_\text{new}(i-1)$ denoting the respective number of persistent and newborn particles at the previous time step.

\item \emph{Predicting:}
As in the \ac{phd-dpf}, each particle is propagated through the system model according to 
\begin{align}
\boldsymbol{s}^p_\text{pers}(i) = \boldsymbol{F} \boldsymbol{s}^p_\text{tot}(i), \qquad p = 1, \ldots, N_\text{pers}(i) = N_\text{tot}(i)
\label{eq:mspphdf_pred1}
\end{align}
to become a persistent particle. The corresponding weights are multiplied with the probability of survival $p_S$ as
\begin{align}
\begin{aligned}
w^p_\text{pers}(i|i-1) &= p_S w^p_\text{tot}(i), \qquad p = 1, \ldots, N_\text{pers}(i).
\end{aligned}
\label{eq:mspphdf_pred2}
\end{align}

\item \emph{Measuring:} The sensor nodes obtain measurements of the targets.

\item \emph{Pre-Clustering:} Since there might be more than one measurement per target, the measurements of the entire network are pre-clustered before the weighting step and each measurement is assigned a label $C(\boldsymbol{z})$ that reflects the cardinality of its own cluster. This can be done, for instance, using single-linkage clustering \cite{everitt2011cluster}. The clustering is based on the distance between measurements, i.e., spatially close measurements are assumed to stem from the same target. Hence, when two or more targets are too close to each other, cardinality errors may occur.

\item \emph{Weighting:} All available target measurements, which comprise the set $\Sigma_i$, are used to update the persistent particle weights according to 
\begin{align}
w^p_\text{pers}(i) = &\Bigg[1 - p_D +  \sum_{\boldsymbol{z}_j \in \Sigma_i} w^p_{j,\text{update}}(i) \Bigg] w^p_\text{pers}(i|i-1),
\label{eq:ms-pphdf_update}
\end{align}
with
\begin{align}
w^p_{j,\text{update}}(i) &= \frac{p_D f_i(\boldsymbol{z}_j\;|\;\boldsymbol{x}^p(i))}{\left(\lambda_\text{FA} c_\text{FA}(\boldsymbol{z}_j) +  \mathcal{L}(\boldsymbol{z}_j)\right) C(\boldsymbol{z}_j)},
\label{eq:ms-pphdf-update1}
\end{align}
and
\begin{align}
\mathcal{L}(\boldsymbol{z}_j) = \sum_{q = 1}^{N_\text{pers}(i)} p_D f_i(\boldsymbol{z}_j\;|\;\boldsymbol{x}^q(i)) w^q_\text{pers}(i|i-1).
\label{eq:ms-pphdf_update2}
\end{align}
Note that---in contrast to the \ac{phd-dpf}---the weighting step is applied only once using the entire set of measurements. Therefore---and since there might be more than one measurement per target---we have to ensure that the weight update terms $w^p_{j,\text{update}}$---and consequently the particle weights---still sum to the number of targets present. This is done by normalizing Equation~(\ref{eq:ms-pphdf-update1}) with $C(\boldsymbol{z}_j)$, i.e., the cardinality of the cluster to which the current measurement $\boldsymbol{z}_j$ belongs.

Afterwards, we form the set $\Sigma_{i,\text{cand}}$ of candidate measurements for the ATB step according to
\begin{align}
\begin{aligned}
	\Sigma_{i,\text{cand}} = \Sigma_i \backslash \Big\{\boldsymbol{z}_{m_p}\;\Big|\;m_p &= \arg \underset{j}{\max}\ w^p_{j,\text{update}}(i), p = 1,...,N_\text{pers}(i)\Big\}.
	\end{aligned}
\end{align}

\item \emph{Resampling:} The expected number of targets $\hat{N}_\text{tgt}(i)$ is calculated from the total persistent particle mass as
\begin{align}
\hat{N}_\text{tgt}(i) = \left\lfloor \sum_{p = 1}^{N_\text{tot}(i)} w^p_\text{pers}(i) \right\rceil.
\label{eq:ms_ntgt}
\end{align}
Consequently, the number of persistent particles is updated according to
\begin{align}
N_\text{pers}(i) = \hat{N}_\text{tgt}(i) N_P.
\end{align}
Furthermore, the set of persistent particles is resampled by drawing $N_\text{pers}(i)$ particles with probability $\frac{w^p_\text{pers}(i)}{\hat{N}_\text{tgt}(i)}$. Then, the weights are reset to equal values as
\begin{align}
w^p_\text{pers}(i) = \frac{\hat{N}_\text{tgt}(i)}{N_\text{pers}(i)}, \qquad p = 1, \ldots, N_\text{pers}(i).
\label{eq:ms_w_res}
\end{align}

\item \emph{Clustering:} In contrast to the \ac{phd-dpf}, there is only one estimate of the expected number of targets. Hence, we can use \emph{$k$-means clustering} \cite{macqueen1967some} to find the estimated target states by grouping the resampled particles into $\hat{N}_\text{tgt}(i)$ clusters and calculating the centroid of each cluster.

\item \emph{Roughening:} Roughening is performed analogously to the \ac{phd-dpf}.
\item \emph{Adaptive Target Birth:} $N_P$ new particles are placed randomly around each candidate measurement $\boldsymbol{z}_j \in \Sigma_{i,\text{cand}}$ yielding a total number of $N_\text{new}(i) = N_P\cdot|\Sigma_{i,\text{cand}}|$ newborn particles. The corresponding weights are chosen according to
\begin{align}
w^p_\text{new}(i) = \frac{p_B}{N_\text{new}(i)}, \qquad p = 1, \ldots, N_\text{new}(i),
\end{align}
where $p_B$ is the probability of birth.
\end{itemize}

The pseudo-code of the \ac{ms-pphdf} is given in Table \ref{alg:ms-pphdf}.

\begin{table*}[p]
  \centering
  \begin{framed}
    \begin{algorithmic}[1]
      \STATE \textbf{input:} $d, E_c, K, n, N, N_P, p_B, p_S, \lambda_\text{FA}, c_\text{FA}$ 
      \STATE \textbf{initialize:} $\left\{\boldsymbol{s}^p_{\text{coll}}(0),w^p_{\text{coll}}(0)\right\}_{p=1}^{N_{\text{coll}}(0)} =\left\{\boldsymbol{s}^p_{\text{new}}(0),w^p_{\text{new}}(0)\right\}_{p=1}^{N_{\text{new}}(0)} = \emptyset $.
      \WHILE{$i\le n$}
        \STATE Merge the sets of persistent and newborn particles with corresponding weights:
        \begin{equation*}
        	\left\{\boldsymbol{s}^p_{\text{tot}}(i),w^p_{\text{tot}}(i)\right\}_{p=1}^{N_{\text{tot}}(i)} = \left\{\boldsymbol{s}^p_{\text{pers}}(i-1),w^p_{\text{pers}}(i-1)\right\}_{p=1}^{N_{\text{pers}}(i-1)} \cup \left\{\boldsymbol{s}^p_{\text{new}}(i-1),w^p_{\text{new}}(i-1)\right\}_{p=1}^{N_{\text{new}}(i-1)}.
        \end{equation*}
		\FOR{$p=1,\ldots,N_{\text{tot}}(i)$}
			\STATE Predict the new state of each particle and update the weight with the probability of survival $p_S$:
			\begin{align*}
\boldsymbol{s}^p_{\text{pers}}(i) &= \boldsymbol{F} \boldsymbol{s}^p_{\text{tot}}(i)\\
w^p_{\text{pers}}(i|i-1) &= p_S w^p_{\text{tot}}(i).
\end{align*}
			\STATE Cluster the measurements using single-linkage clustering and assign each measurement $\boldsymbol{z}$ a label $C(\boldsymbol{z})$ reflecting the cardinality of its cluster.
			\STATE Update the weights using the measurements of the entire network:

\begin{align*}
w^p_\text{pers}(i) = &\Bigg[1 - p_D +  \sum_{\boldsymbol{z}_j \in \Sigma_i} w^p_{j,\text{update}}(i) \Bigg] w^p_\text{pers}(i|i-1)\\
w^p_{j,\text{update}}(i) &= \frac{p_D f_i(\boldsymbol{z}_j\;|\;\boldsymbol{x}^p(i))}{\left(\lambda_\text{FA} c_\text{FA}(\boldsymbol{z}_j) +  \mathcal{L}(\boldsymbol{z}_j)\right) C(\boldsymbol{z}_j)},\\
\mathcal{L}(\boldsymbol{z}_j) &= \sum_{q = 1}^{N_\text{pers}(i)} p_D f_i(\boldsymbol{z}_j\;|\;\boldsymbol{x}^q(i)) w^q_\text{pers}(i|i-1).
\end{align*}
		\ENDFOR
		\STATE Form the set of candidate measurements for ATB:
		\begin{align*}
	\Sigma_{i,\text{cand}} = \Sigma_i \backslash \Big\{\boldsymbol{z}_{m_p}\;\Big|\;m_p &= \arg \underset{j}{\max}\ w^p_{j,\text{update}}(i),\quad p = 1,...,N_{\text{pers}}(i), \quad j = 1,...,\left|\Sigma_i\right|\Big\}.
\end{align*}
		\STATE Calculate the estimated number of targets:
		\begin{align*}
\hat{N}_{\text{tgt}}(i) = \left\lfloor \sum_{p = 1}^{N_{\text{tot}}(i)} w^p_{\text{pers}}(i) \right\rceil.
\end{align*}
        \STATE Resample $\hat{N}_{\text{pers}}(i) = \hat{N}_{\text{tgt}}(i)N_P$ particles and reset the weights:
        \begin{align*}
w^p_{\text{pers}}(i) = \frac{\hat{N}_{\text{tgt}}(i)}{N_{\text{pers}}(i)}, \qquad p = 1, \ldots, N_{\text{pers}}(i).
\end{align*}
        \STATE Find the set of estimated target states $\left\{\hat{\boldsymbol{s}}^l(i)\right\}_{l=1}^{\hat{N}_{\text{tgt}}(i)}$ by using \mbox{$k$-means} clustering and calculating the centroid of each cluster.
        \STATE Add an independent jitter to each particle using a component-wise standard deviation of:
        \begin{align*}
\sigma_c^j(i) = K E_c N_{\text{pers}}(i)^{-1/d}.
\end{align*}

        \STATE Place $N_P$ new particles randomly around each candidate measurement $\boldsymbol{z}_j \in \Sigma_{i,\text{cand}}$. Set the weights as:
        \begin{align*}
w^p_{\text{new}}(i) = \frac{p_B}{N_{\text{new}}(i)}, \qquad p = 1, \ldots, N_{\text{new}}(i).
\end{align*}
      \STATE $i\gets i+1$ 
      \ENDWHILE
      \RETURN 
    \end{algorithmic}
  \end{framed}
  \caption{The Multi-Sensor Particle PHD Filter.}
  \label{alg:ms-pphdf} 
\end{table*}

\subsection{Computational Complexity and Communication Load}
In this section we analyze the computational complexity and the communication load of the \ac{ms-pphdf}. The following steps are performed at every time instant $i$ but time dependency is omitted for simplicity:
\begin{itemize}
	\item \emph{Prediction}: The prediction step described by Equations \eqref{eq:mspphdf_pred1} and \eqref{eq:mspphdf_pred2} is performed for each of the $N_\text{tot}$ particles and is linear in the dimensionality $d$.\\
	$\Rightarrow \mathcal{O}(N_\text{tot}d)$
	\item \emph{Pre-Clustering}: The pre-clustering step relies on single-linkage clustering. The complexity is therefore cubic in the total number of measurements $N_\text{meas}$. \cite{murtagh1983survey}\\
	$\Rightarrow \mathcal{O}(N_\text{meas}^3)$
	\item \emph{Weighting}: Each particle is updated in the weighting step given by Equations \eqref{eq:ms-pphdf_update}-\eqref{eq:ms-pphdf_update2}. The weight update as well as the designation of candidate measurements for ATB depends on  the number of measurements $N_\text{meas} = \left|\Sigma\right|$.\\
	$\Rightarrow \mathcal{O}(N_\text{tot}N_\text{meas})$ 
	\item \emph{Resampling}: The estimation of the number of targets and the resampling step in Equations \eqref{eq:ms_ntgt}-\eqref{eq:ms_w_res} are linear in the number of particles used for the calculation \cite{gustafsson2010particle}.\\
	$\Rightarrow \mathcal{O}(N_\text{tot} + N_\text{tgt}N_P)$
	\item \emph{Clustering}: In contrast to the \ac{phd-dpf} we can use \mbox{$k$-means} clustering. The complexity of Lloyd's implementation is given by \cite{inaba1994applications}\\ 
	$\Rightarrow \mathcal{O}((N_\text{tgt}N_P)^{dN_\text{tgt}+1}\log(N_\text{tgt}N_P))$.
	\item \emph{Roughening}: Roughening is linear in the dimensionality of the particles and their number.\\
	$\Rightarrow \mathcal{O}(N_\text{tgt}N_Pd)$
	\item \emph{Adaptive Target Birth}: The birth process depends on the number of particles per target $N_P$ as well as the number of candidate measurements $N_\text{cand}$.\\
	$\Rightarrow \mathcal{O}(N_PN_\text{cand})$
	\end{itemize}
	
	In summary, the computational complexity of the \ac{ms-pphdf} is largely comparable to that of the \ac{phd-dpf}. The only exception is the pre-custering step, which scales cubicly with the total number of measurements and adds additional complexity to the algorithm. As a tradeoff the communication load of the \ac{ms-pphdf} clearly is lower compared to the \ac{phd-dpf} because there is only the initial transmission of measurements from the nodes to the fusion center. However, considering a setup with relatively small communication radii, this initial communication step requires a lot of relaying and leads to high traffic density in the vicinity of the fusion center. Furthermore, this communication structure exhibits a single point of failure while a distributed sensor network is inherently redundant.

\section{Simulations}
In this section, we evaluate the performance of the \ac{phd-dpf} as well as the \ac{ms-pphdf} for tracking multiple targets in a sensor network with 1-coverage. To this end, we consider Gaussian measurement noise of different variance as well as $\varepsilon$-contaminated noise with different contamination ratios to investigate the robustness of the algorithms. In addition, the performance for different amounts of clutter is analyzed. We compare the proposed algorithms to the alternative distributed \ac{phd-pf} from \cite{Uney2010}, which will be referred to as \ac{ddf-pphdf}. Here, each node runs its own \ac{phd-pf} using only its own measurements. In a subsequent step, the particles are distributed over the neighborhood and reweighted by fusing their corresponding Exponential Mixture Densities.

Furthermore, we formulate the \acs{dpcrlb} as a lower bound for evaluating the performance of the three algorithms in terms of the \acs{ospa} \cite{Schuhmacher2008} metric. In our simulations, we compute the \acs{ospa} metric with respect to the joint set of target state estimates of the entire active network. The latter is found by clustering the target state estimates of all active nodes. Furthermore, we consider the squared \acs{ospa} metric scaled by the number of targets, i.e., $N_\text{tgt} \cdot \left(\bar{d}_p^{(c)}\right)^2$, as in \cite{braca2013asymptotic}. That way, we can use the \ac{dpcrlb}, which will be introduced in the following, as a benchmark. 


 \subsection{The Distributed Posterior Cram\'{e}r-Rao Lower Bound (DPCRLB)}
Rather than evaluating the performance of the different \ac{mtt} algorithms based on an error metric, it makes more sense to derive a minimum variance bound on the estimation error, which enables an absolute performance evaluation. For time-invariant statistical models, the most commonly used bound is the \ac{crlb}, which is given by the inverse of Fisher's information matrix \cite{kay1993fundamentals}. In \cite{braca2013asymptotic} and \cite{Braca2012}, the \ac{crlb} is used in the context of multi-sensor \ac{mtt} of an unknown number of unlabeled targets in order to evaluate the performance, as well as prove the asymptotic efficiency of the PHD as the number of nodes goes to infinity. Since we are more interested in the tracking behavior of a fixed network over time, we resort to the \ac{pcrlb}, which is an extension of the \ac{crlb} for the time-variant case \cite{van2004detection}. This bound can be calculated sequentially with the help of a Riccati-like recursion derived in \cite{Tichavsky1998}. Furthermore, in \cite{Hue2002} and \cite{Hue2002b}, the \ac{pcrlb} is adapted for an \ac{mtt} scenario in which the tracker can obtain more than one measurement per target. 

Let $\pi_i^m, m=1, ...,M$ denote the probability that any measurement is associated with target $m$ at time instant $i$ as defined in \cite{Hue2002a}. With the corresponding stochastic process $\Pi_i^m$, the new stochastic process of association probabilities and target states to be estimated becomes $\boldsymbol{\Phi}_i = \left( \Pi_i^{1:M}, \boldsymbol{\Xi}_i^{1:M} \right)$. Fisher's information matrix $J_{\boldsymbol{\Phi}_{i}} = \begin{bmatrix}J_{\Pi_{i}} & J_{\Pi_{i}}^{\boldsymbol{\Xi}_i} \\ J^{\Pi_{i}}_{\boldsymbol{\Xi}_i} & J_{\boldsymbol{\Xi}_{i}}\end{bmatrix}$ can now be formed as described in \cite{Hue2002b} and \cite{Hue2002}. However, as the number of targets varies over time, i.e., targets might enter or exit the \ac{roi}, $J_{\boldsymbol{\Phi}_{i}}$ has to be expanded or shrunk in the inverse matrix domain as described in \cite{Bessell2003}. The \ac{pcrlb} $B_i$ at time instant $i$ can be obtained as the trace of the inverted submatrix $J_{\boldsymbol{\Xi}_{i}}$ according to \cite{Hue2002b}

\begin{align}
B_i = \text{trace}\left\{\left[ J_{\boldsymbol{\Xi}_{i}} - J^{\Pi_{i}}_{\boldsymbol{\Xi}_i} J_{\Pi_{i}}^{-1} J_{\Pi_{i}}^{\boldsymbol{\Xi}_i}\right]^{-1}\right\}.
\end{align}

Note that, in a distributed \ac{mtt} scenario, $B_i$ corresponds to a lower bound on the estimation error of a central processing unit with access to all measurements. Since we are interested in completely distributed \ac{mtt} with in-network processing, we extend the \ac{pcrlb} to the \acl{dpcrlb} (\acs{dpcrlb}). To this end, each node $k$ computes its own \ac{pcrlb} $B_i^k$ considering only the measurements of its two-hop neighborhood, which is given by 
\begin{align}
\mathcal{N}_k^{(2)} = \bigcup_{l \in \mathcal{N}_k}\mathcal{N}_l,
\end{align}
i.e., the neighbors of node $k$ and their neighbors. Furthermore, only the targets within the sensing range of $\mathcal{N}_k$ are taken into account. Clearly, only nodes with a neighborhood in the vicinity of at least one target will be able to calculate a \ac{pcrlb}. The \ac{dpcrlb} $B_{i,\text{dist}}$ at time instant $i$ is then obtained by averaging over these values according to


\begin{align}
B_{i,\text{dist}} = \frac{1}{|\mathcal{M}|} \sum_{k \in \mathcal{M}} B_i^k,
\end{align}
where $\mathcal{M}$ is the set of all nodes that are able to compute a \ac{pcrlb}.

\subsection{Simulation Setup}
In the following simulations, a static sensor network as depicted in Figure~\ref{fig:mtt_tracks}) is used to perform \ac{mtt}. The network is centered around the point of origin $\left[0,0\right]^\top$ and distributed such that 1-coverage of the \ac{roi} is guaranteed. It covers an area of approximately $2500~\text{m}^2$. Clutter is assumed Poisson and uniformly distributed over the sensing range of each node with an average rate of $\lambda_\text{FA} = 0.1$ and $0.3$. Moreover, we consider Gaussian measurement noise with variance $\sigma_r^2 = 0.1$ and 0.3 as well as $\varepsilon$-contamination noise with a ten-times higher variance and a contamination rate of $\varepsilon=0.1$ and 0.3. For the sake of simplicity, collisions between targets and sensor nodes are neglected. 

An overview of all simulation parameters is given in Table~\ref{tab:ov_sim}. Since the purpose of this work is to introduce the \ac{ms-pphdf} as well as the \ac{phd-dpf}, verify their functionality, and compare them to alternative approaches, we consider a rather simple scenario with a high probability of detection and relatively low clutter levels. In our future work, we will study more sophisticated scenarios to define possible breakdown points of our algorithms.

\begin{table}[!t]
\centering
\begin{tabular}{lll}
\toprule
\textsc{Variable} & \textsc{Value} & \textsc{Description} \\
\midrule

 $\Delta i$ & 1~s & time step of the tracking algorithm\\
 $N$ & 30 & number of nodes\\
 \midrule
 $\sigma_r^2$ & 0.1, 0.3~$\text{m}^2$ & componentwise power of meas. noise\\
 $\sigma_q^2$ & 0.01~$\text{m}^2$ & componentwise power of state noise\\
 $\varepsilon$ & 0.1, 0.3 & contamination ratio\\
 \midrule

 $R_\text{com}$ & $2 R_\text{sen}$ & communication radius\\
 $R_\text{sen}$ & 6~m & sensing radius\\
\midrule
$E_c$ & 6 & empirical interval length for jitter\\
 $K$ & 0.2 & tuning constant for roughening\\
 $N_P$ & 500 & number of particles per target\\
 $p_B$ & 0.8 & probability of birth\\
 $p_D$ & 0.95 & probability of detection\\
 $p_S$ & 0.98 & probability of survival\\
 \midrule
 $\lambda_\text{FA}$ & 0.1, 0.3 & average no. of false alarms / clutter\\
 $c_\text{FA}(\boldsymbol{z})$ & $\frac{1}{\pi R_\text{sen}^2}$ & PDF of false alarms / clutter (uniform)\\
  \midrule
$c$ & 2 & cut-off value (OSPA)\\
$p$ & 2 & order of the OSPA metric\\
\bottomrule

\end{tabular}
\caption{Simulation parameters}
\label{tab:ov_sim}
\end{table}

We use the \ac{ms-pphdf}, the \ac{phd-dpf}, as well as the \ac{ddf-pphdf} to track three targets for $i = 0, ..., 30$. The targets enter the \ac{roi} at time steps $i = {0, 9, 14}$ from the north, south, and west, respectively. A Monte Carlo simulation with $N_\text{MC} = 1000$ runs is performed to evaluate the performance of the tracking algorithms in terms of the \ac{ospa} metric. Note that the target trajectories as shown in Figure~\ref{fig:mtt_tracks}) are deterministic, as is often the case in target tracking simulations \cite{Bessell2003} in order to guarantee the comparability of the different Monte Carlo runs regarding, for instance, the number of targets present.

\begin{figure*}[!t]
\psfrag{atbPHD-PF}[cl][cl][0.8]{\footnotesize MS-PPHDF}
\psfrag{atbdiffPHD-PF}[cl][cl][0.8]{\footnotesize D-PPHDF}
\psfrag{distPHD-PF}[cl][cl][0.8]{\footnotesize DDF-PPHDF}
\psfrag{PCRLB}[cl][cl][0.8]{\footnotesize DPCRLB}
\psfrag{true}[cl][cl][0.8]{\footnotesize true number}
\psfrag{y}[bc][cc][0.9]{\footnotesize $N_\text{tgt}\cdot\left(\bar{d}_2^{(2)}\right)^2$ and $B_{i,\text{dist}}$}
\psfrag{x}[tc][tc][0.9]{\footnotesize Time steps $i$ (seconds)}
\psfrag{0}[Bc][Bc][0.7]{\footnotesize 0}
\psfrag{5}[cc][cc][0.7]{\footnotesize 5}
\psfrag{15}[cc][cc][0.7]{\footnotesize 15}
\psfrag{25}[cc][cc][0.7]{\footnotesize 25}
\psfrag{10}[cc][cc][0.7]{\footnotesize 10}
\psfrag{20}[cc][cc][0.7]{\footnotesize 20}
\psfrag{30}[cc][cc][0.7]{\footnotesize 30}
\psfrag{35}[cc][cc][0.7]{\footnotesize 35}
\psfrag{0.2}[cc][cc][0.7]{\footnotesize 0.2}
\psfrag{0.4}[cc][cc][0.7]{\footnotesize 0.4}
\psfrag{0.6}[cc][cc][0.7]{\footnotesize 0.6}
\psfrag{0.8}[cc][cc][0.7]{\footnotesize 0.8}
\psfrag{1}[cc][cc][0.7]{\footnotesize 1}
\psfrag{1.2}[cc][cc][0.7]{\footnotesize 1.2}
\psfrag{1.4}[cc][cc][0.7]{\footnotesize 1.4}
\psfrag{1.6}[cc][cc][0.7]{\footnotesize 1.6}
\psfrag{1.8}[cc][cc][0.7]{\footnotesize 1.8}
\psfrag{2}[cc][cc][0.7]{\footnotesize 2}
\centering

\subfloat[Squared and scaled OSPA compared to DPCRLB, $\sigma_r^2 = 0.1$]{
\includegraphics[width=.45\textwidth]{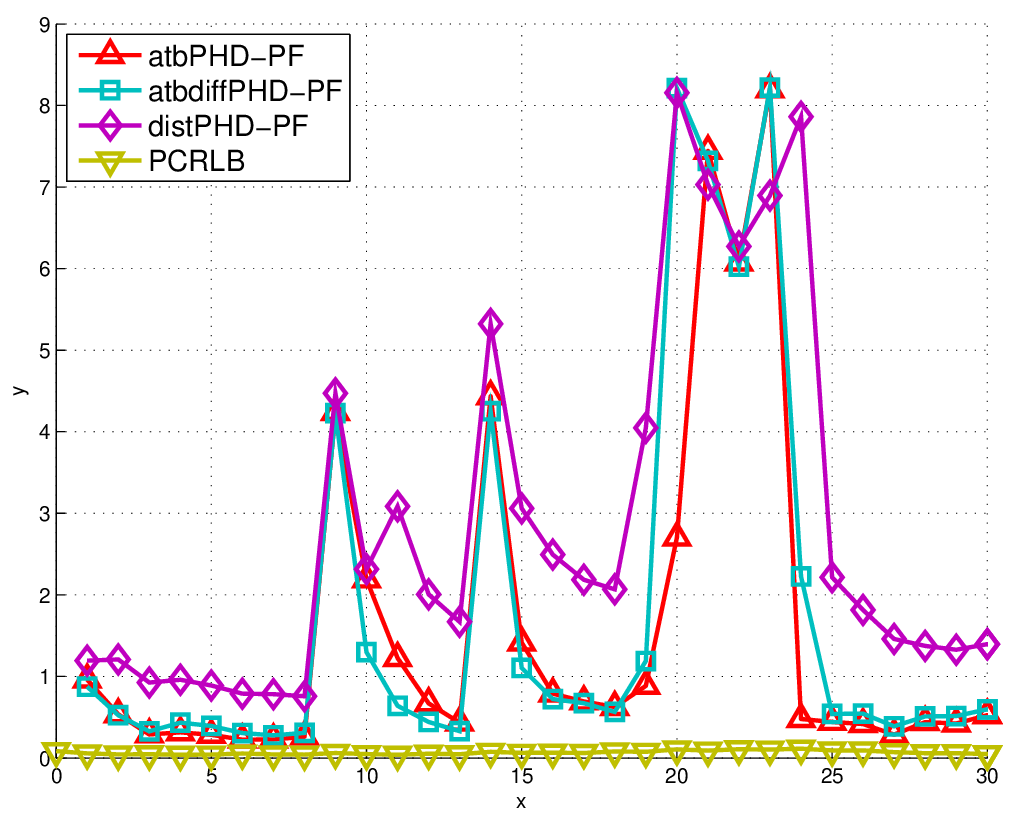}
\label{fig:ospa01x}
}\qquad
\subfloat[Estimated number of targets, $\sigma_r^2 = 0.1$]{
\psfrag{y}[bc][cc][0.9]{\footnotesize $N_\text{est}$}
\includegraphics[width=.45\textwidth]{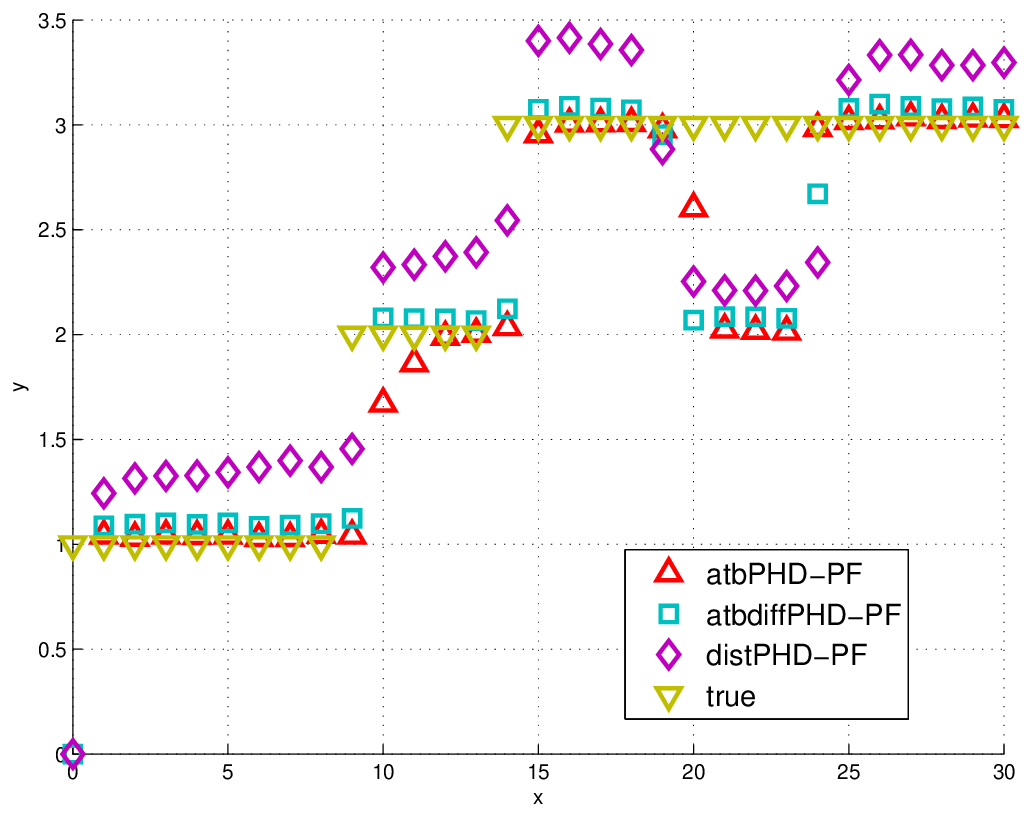}
\label{fig:nest01x}
}\\
\subfloat[Squared and scaled OSPA compared to DPCRLB, $\sigma_r^2 = 0.3$]{
\includegraphics[width=.45\textwidth]{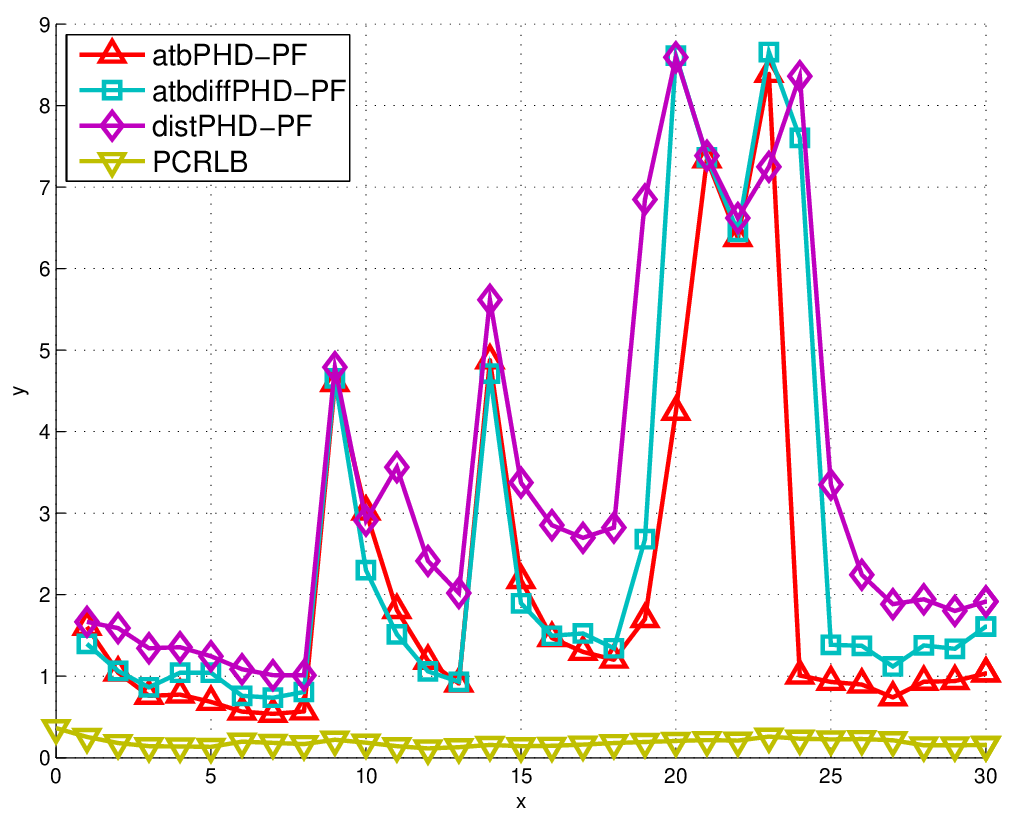}
\label{fig:ospa03x}
}\qquad
\subfloat[Estimated number of targets, $\sigma_r^2 = 0.3$]{
\psfrag{y}[bc][cc][0.9]{\footnotesize $N_\text{est}$}
\includegraphics[width=.45\textwidth]{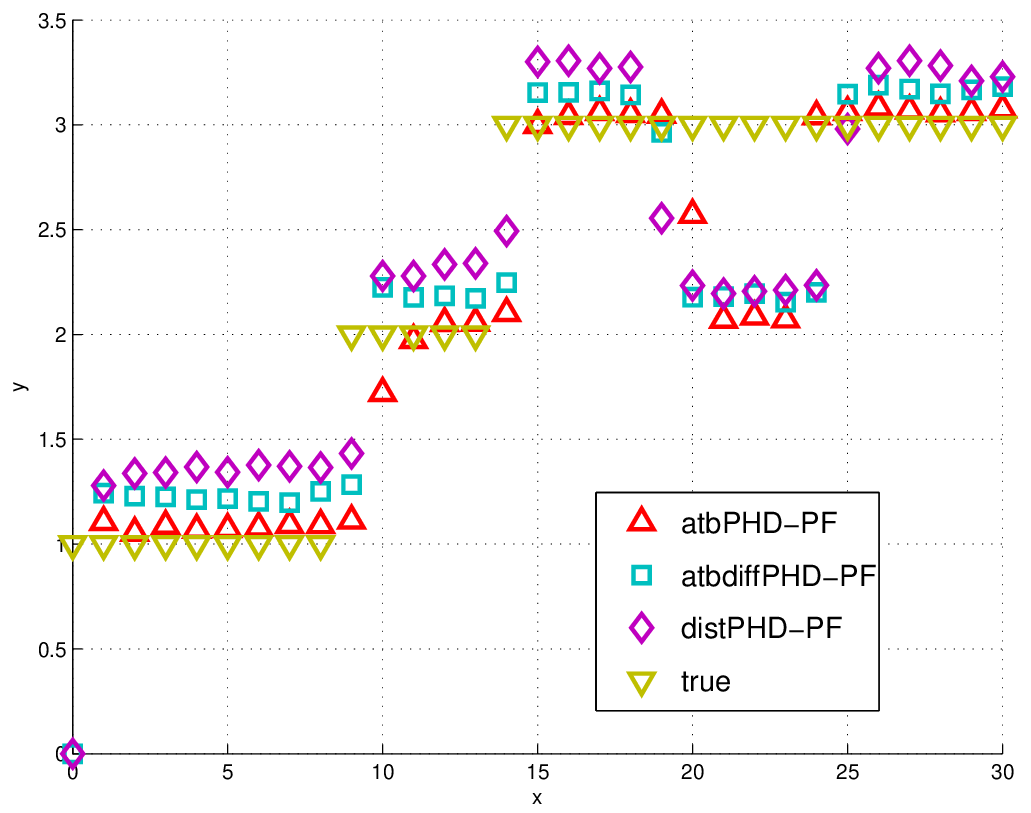}
\label{fig:nest03x}
}
\caption{Simulation I: Results for Gaussian noise and clutter rate $\lambda_\text{FA}=0.1$. The left part of the figure shows the squared and scaled \acl{ospa} (\acs{ospa}) metric for each algorithm compared to the \acl{dpcrlb} (\acs{dpcrlb}), while the right part compares the estimated to the true number of targets.}
\label{fig:simresults}
\end{figure*}

\begin{figure*}[!t]
\psfrag{atbPHD-PF}[cl][cl][0.8]{\footnotesize MS-PPHDF}
\psfrag{atbdiffPHD-PF}[cl][cl][0.8]{\footnotesize D-PPHDF}
\psfrag{distPHD-PF}[cl][cl][0.8]{\footnotesize DDF-PPHDF}
\psfrag{PCRLB}[cl][cl][0.8]{\footnotesize DPCRLB}
\psfrag{true}[cl][cl][0.8]{\footnotesize true number}
\psfrag{y}[bc][cc][0.9]{\footnotesize $N_\text{tgt}\cdot\left(\bar{d}_2^{(2)}\right)^2$ and $B_{i,\text{dist}}$}
\psfrag{x}[tc][tc][0.9]{\footnotesize Time steps $i$ (s)}
\psfrag{0}[Bc][Bc][0.7]{\footnotesize 0}
\psfrag{5}[cc][cc][0.7]{\footnotesize 5}
\psfrag{15}[cc][cc][0.7]{\footnotesize 15}
\psfrag{25}[cc][cc][0.7]{\footnotesize 25}
\psfrag{10}[cc][cc][0.7]{\footnotesize 10}
\psfrag{20}[cc][cc][0.7]{\footnotesize 20}
\psfrag{30}[cc][cc][0.7]{\footnotesize 30}
\psfrag{35}[cc][cc][0.7]{\footnotesize 35}
\psfrag{0.2}[cc][cc][0.7]{\footnotesize 0.2}
\psfrag{0.4}[cc][cc][0.7]{\footnotesize 0.4}
\psfrag{0.6}[cc][cc][0.7]{\footnotesize 0.6}
\psfrag{0.8}[cc][cc][0.7]{\footnotesize 0.8}
\psfrag{1}[cc][cc][0.7]{\footnotesize 1}
\psfrag{1.2}[cc][cc][0.7]{\footnotesize 1.2}
\psfrag{1.4}[cc][cc][0.7]{\footnotesize 1.4}
\psfrag{1.6}[cc][cc][0.7]{\footnotesize 1.6}
\psfrag{1.8}[cc][cc][0.7]{\footnotesize 1.8}
\psfrag{2}[cc][cc][0.7]{\footnotesize 2}

\centering

\subfloat[Squared and scaled OSPA compared to DPCRLB, $\sigma_r^2 = 0.1$]{
\includegraphics[width=.45\textwidth]{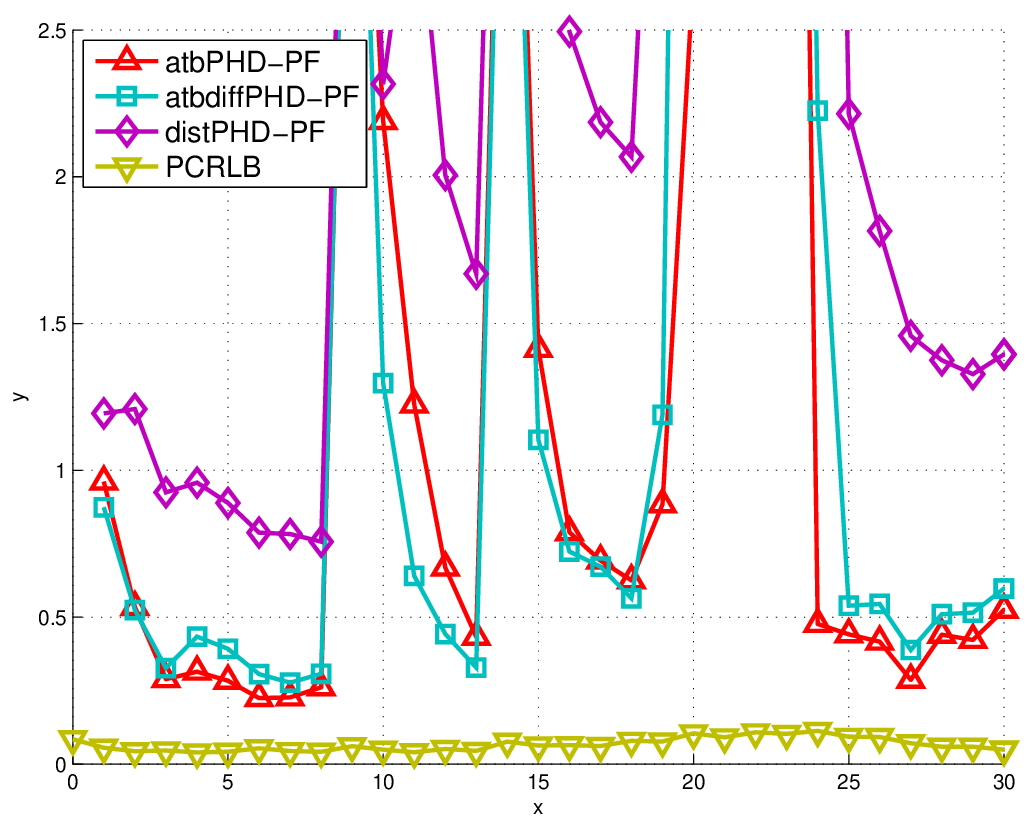}
\label{fig:ospa01xzoom}
}\qquad
\subfloat[Squared and scaled OSPA compared to DPCRLB , $\sigma_r^2 = 0.3$]{
\includegraphics[width=.45\textwidth]{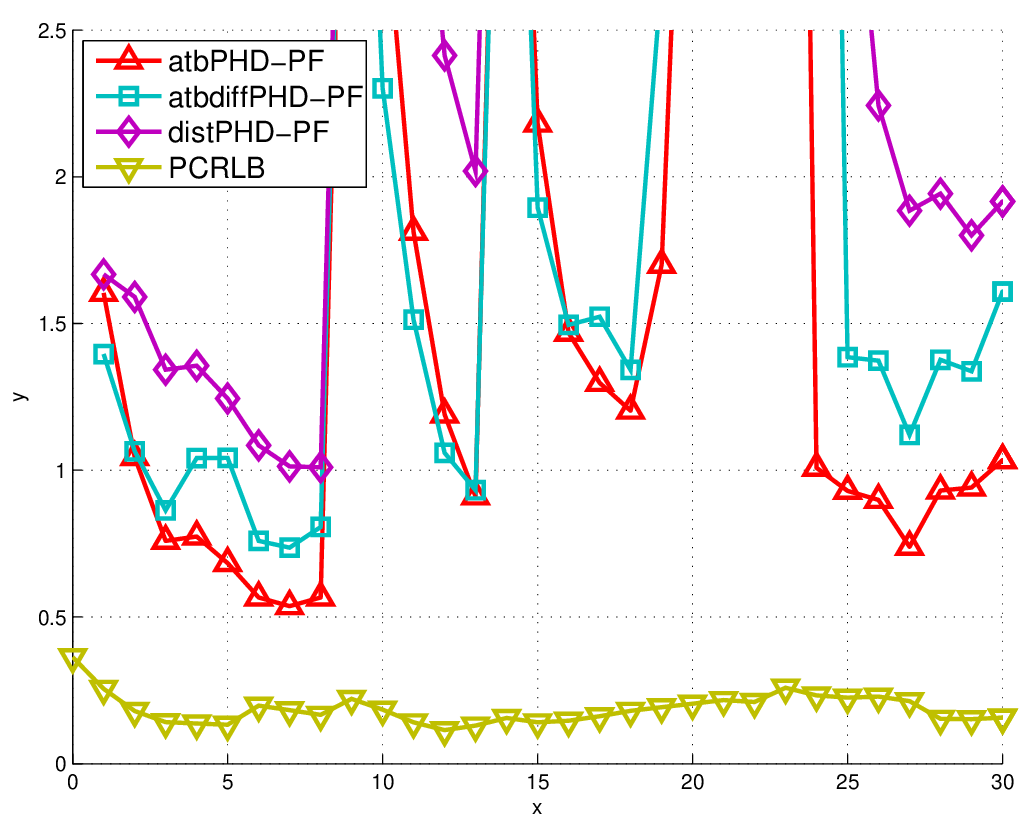}
\label{fig:ospa03xzoom}
}\caption{Simulation I: Results for Gaussian noise and clutter rate $\lambda_\text{FA}=0.3$ (zoomed in). The squared and scaled \acl{ospa} (\acs{ospa}) metrics of using the \acl{ms-pphdf} (\acs{ms-pphdf}), the \acl{phd-dpf} (\acs{phd-dpf}), and the \acl{ddf-pphdf} (\acs{ddf-pphdf}), are compared to the \acl{dpcrlb} (\acs{dpcrlb}).}
\label{fig:simresultszoom}
\end{figure*}

\subsection{Simulation I: Results }
In the first simulation, we compare the performance of the MS-PPHDF and the D-PPHDF to the alternative DDF-PPHDF and the \ac{dpcrlb}, which serves as a benchmark. Measurement noise is zero-mean Gaussian with variance $\sigma_r^2=0.1, 0.3$ and the average number of clutter is $0.1$.

The simulation results are depicted in Figure \ref{fig:simresults}. While the top part considers zero-mean Gaussian measurement noise with a per-component variance of $\sigma_r^2 = 0.1$, the bottom part shows the results for $\sigma_r^2 = 0.3$. In addition to evaluating the performance of the \ac{ms-pphdf}, the \ac{phd-dpf}, and the \ac{ddf-pphdf} in terms of the squared and scaled \ac{ospa} metric over time and comparing it to the \ac{dpcrlb} as can be seen in the left part of the figure, we also look at the estimated number of targets, which is depicted in the right column. Since the \ac{ospa} metric contains a penalty for an erroneous estimate of the number of targets, this side-by-side comparison facilitates the interpretation of the tracking results.

Let us start by considering Figures \ref{fig:ospa01x}) and \ref{fig:nest01x}), i.e., the case of $\sigma_r^2 = 0.1$. First of all, we observe that neither tracking algorithm provides an \ac{ospa} value or an estimate of the number of targets for $i = 0$. This is expected and due to \ac{atb}, which initializes new particle clouds based on the measurements from the previous time step. Thus, target birth is delayed by one time step and tracking can only be performed for $i > 0$. The same effect can be witnessed at $i = 9$ and $i = 14$, respectively, which mark the time instants at which targets 2 and 3 enter the \ac{roi}. Here, the \ac{ospa} curves of all trackers exhibit a spike, which is due to the fact that the newborn particles are not yet considered in the tracker and, hence, the number of estimated targets is too low, as can be seen in Figure \ref{fig:nest01x}).

Another sudden rise of all the \ac{ospa} curves can be observed in the time interval $20 \leq i \leq 24$ with a valley at $i = 22$. Looking at the estimated number of targets, we can attribute this phenomenon to the fact that only two of the three targets are recognized by the tracking algorithms. Since the target trajectories are deterministic, we know that in the given time interval targets 2 and 3 cross paths. Due to the inability of the clustering algorithm to separate strongly overlapping sets of measurements, the two targets merge into one as long as they are close to each other. When the two targets occupy almost exactly the same position, i.e., at $i = 22$, the \ac{ospa} metric decreases due to the decrease in measurement variance. As the targets drift apart, the variance and with it the \ac{ospa} metric increases up to the point where the two targets can be recognized as separate again and the corresponding penalty is switched off.

Looking at the overall picture in Figure \ref{fig:ospa01x}), which shows the case of $\sigma_r^2 = 0.1$, it is evident that the centralized \ac{ms-pphdf} and the distributed \ac{phd-dpf} achieve approximately the same performance with \ac{ospa} values closely approaching the \ac{dpcrlb} when the number of targets stays constant. Furthermore, both algorithms deliver very accurate estimates of the number of targets, given they are separable by clustering, as can be seen in Figure \ref{fig:nest01x}). The \ac{ddf-pphdf}, however, continuously exhibits a worse performance than the \ac{phd-dpf}, both in terms of the \ac{ospa} metric as well as the estimated number of targets. This is where the additional communication in the proposed \ac{phd-dpf} shows its strength in reducing uncertainty due to measurement noise and clutter. Apart from achieving worse tracking results, the \ac{ddf-pphdf} also has more difficulty in separating targets 1 and 2 when they cross paths, resulting in an earlier rise and a later fall of the \ac{ospa} metric, compared to our approach.

In the case of $\sigma_r^2 = 0.3$, the overall performance of the different tracking algorithms is very similar to the case of $\sigma_r^2 = 0.1$. In order to make a statement on how the different tracking algorithms compare, let us neglect the penalty due to an erroneous estimate of the number of targets and take a look at Figures \ref{fig:ospa01xzoom}) and \ref{fig:ospa03xzoom}), which are zoomed-in versions of Figures \ref{fig:ospa01x}) and \ref{fig:ospa03x}), respectively.

In Figures \ref{fig:ospa01xzoom}) and \ref{fig:ospa03xzoom}) the \ac{dpcrlb} is given as a benchmark for tracking performance. One can observe that its value is always smaller or equal to the respective measurement variance. As stated before, the centralized \ac{ms-pphdf} and the distributed \ac{phd-dpf} exhibit very similar performance and deliver better tracking results than the \ac{ddf-pphdf}. While the \ac{ms-pphdf} achieves lower \ac{ospa} values than the \ac{phd-dpf} when the number of targets stays constant, i.e., for $3 \leq i \leq 8$ and $24 \leq i \leq 30$, the \ac{phd-dpf} performs better directly after a new target appears, i.e., for $1 \leq i \leq 2$, $10 \leq i \leq 13$, and $15 \leq i \leq 18$. This is likely due to the fact that the two-step communication scheme employed in the \ac{phd-dpf} is able to reduce the impact of measurement noise and clutter faster than the centralized \ac{ms-pphdf} can.

Looking at the case of $\sigma_r^2 = 0.3$ in Figure \ref{fig:ospa03xzoom}), we observe that the higher measurement noise affects the performance of all algorithms, resulting in higher \ac{ospa} curves. While the \ac{ospa} curves of the \ac{ms-pphdf} and the \ac{ddf-pphdf} are proportionally shifted upward by approximately the same value, i.e., they are equally impacted by the higher noise level, the \ac{phd-dpf} seems to be slightly more affected by the change. However it still outperforms the \ac{ddf-pphdf} at all time instants.

All in all, the proposed \ac{phd-dpf} yields better performance than the existing \ac{ddf-pphdf} in estimating the number of targets and tracking them, irrespective of the amount of measurement noise. In addition, it is also a bit faster in delivering correct state estimates of new targets than the centralized \ac{ms-pphdf} and performs only slightly worse once the number of targets stays constant. In our future work, we will look at ways to further improve the performance of the \ac{ms-pphdf} and the \ac{phd-dpf} in order to approach the \ac{dpcrlb} even more closely. 

\begin{figure*}[!t]
\psfrag{atbPHD-PF}[cl][cl][0.8]{\footnotesize MS-PPHDF}
\psfrag{atbdiffPHD-PF}[cl][cl][0.8]{\footnotesize D-PPHDF}
\psfrag{distPHD-PF}[cl][cl][0.8]{\footnotesize DDF-PPHDF}
\psfrag{PCRLB}[cl][cl][0.8]{\footnotesize DPCRLB}
\psfrag{true}[cl][cl][0.8]{\footnotesize true number}
\psfrag{y}[bc][cc][0.9]{\footnotesize $N_\text{tgt}\cdot\left(\bar{d}_2^{(2)}\right)^2$ and $B_{i,\text{dist}}$}
\psfrag{x}[tc][tc][0.9]{\footnotesize Time steps $i$ (seconds)}
\psfrag{0}[Bc][Bc][0.7]{\footnotesize 0}
\psfrag{5}[cc][cc][0.7]{\footnotesize 5}
\psfrag{15}[cc][cc][0.7]{\footnotesize 15}
\psfrag{25}[cc][cc][0.7]{\footnotesize 25}
\psfrag{10}[cc][cc][0.7]{\footnotesize 10}
\psfrag{20}[cc][cc][0.7]{\footnotesize 20}
\psfrag{30}[cc][cc][0.7]{\footnotesize 30}
\psfrag{35}[cc][cc][0.7]{\footnotesize 35}
\psfrag{0.2}[cc][cc][0.7]{\footnotesize 0.2}
\psfrag{0.4}[cc][cc][0.7]{\footnotesize 0.4}
\psfrag{0.6}[cc][cc][0.7]{\footnotesize 0.6}
\psfrag{0.8}[cc][cc][0.7]{\footnotesize 0.8}
\psfrag{1}[cc][cc][0.7]{\footnotesize 1}
\psfrag{1.2}[cc][cc][0.7]{\footnotesize 1.2}
\psfrag{1.4}[cc][cc][0.7]{\footnotesize 1.4}
\psfrag{1.6}[cc][cc][0.7]{\footnotesize 1.6}
\psfrag{1.8}[cc][cc][0.7]{\footnotesize 1.8}
\psfrag{2}[cc][cc][0.7]{\footnotesize 2}
\centering

\includegraphics[width=.45\textwidth]{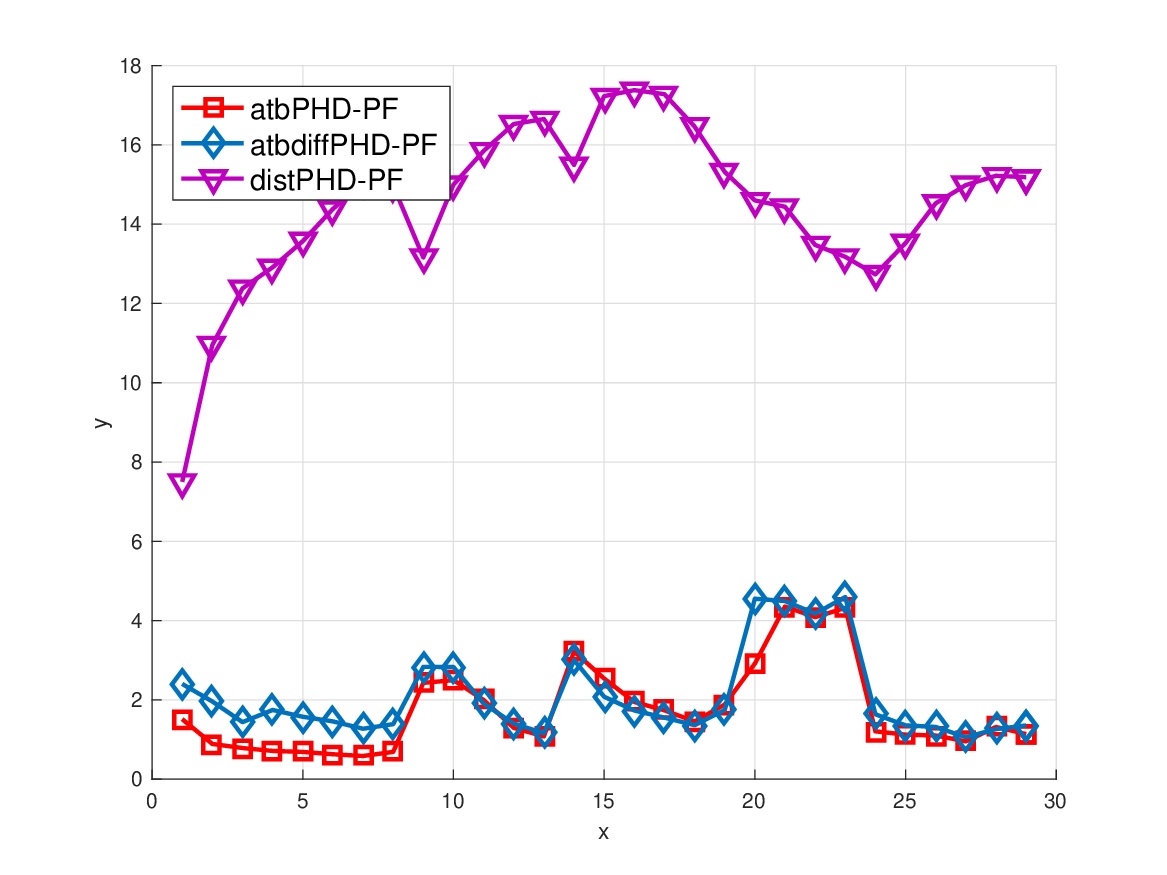}
\qquad
\psfrag{y}[bc][cc][0.9]{\footnotesize $N_\text{est}$}
\includegraphics[width=.45\textwidth]{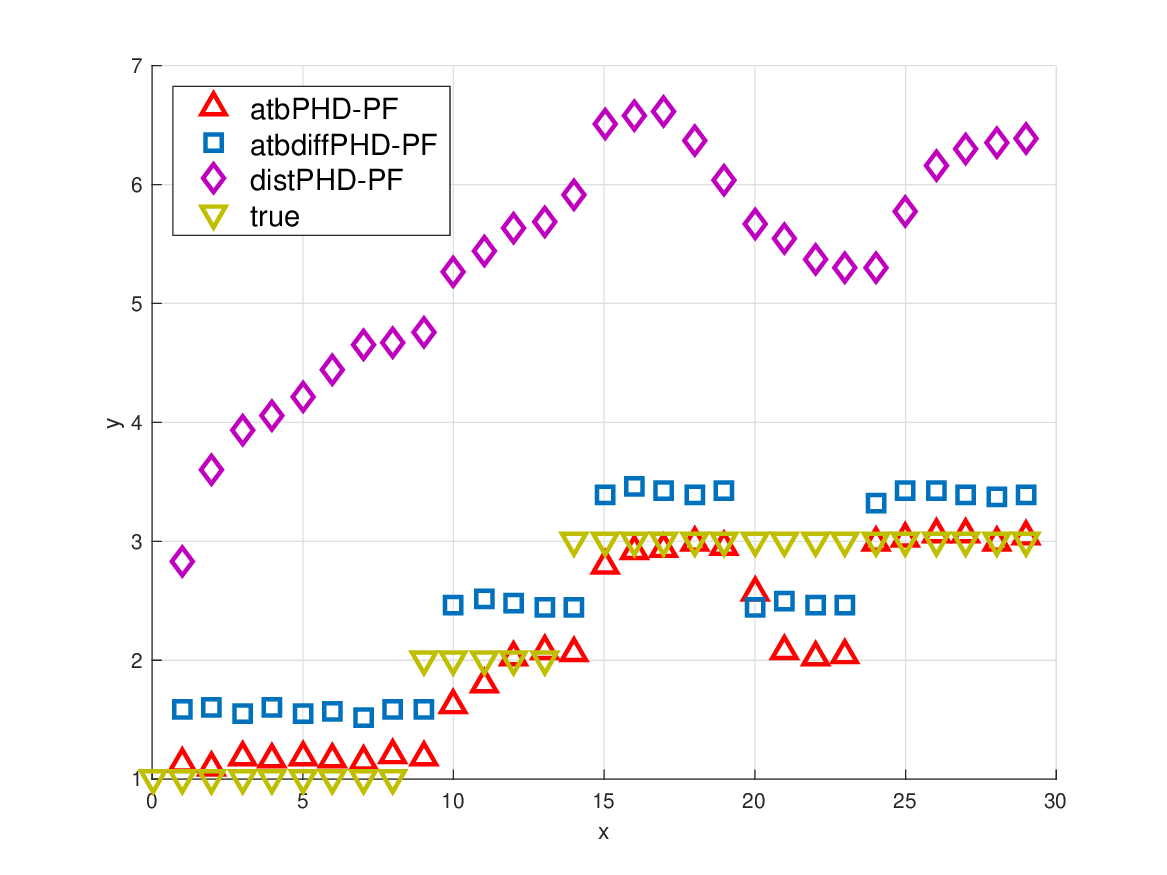}
\\
\includegraphics[width=.45\textwidth]{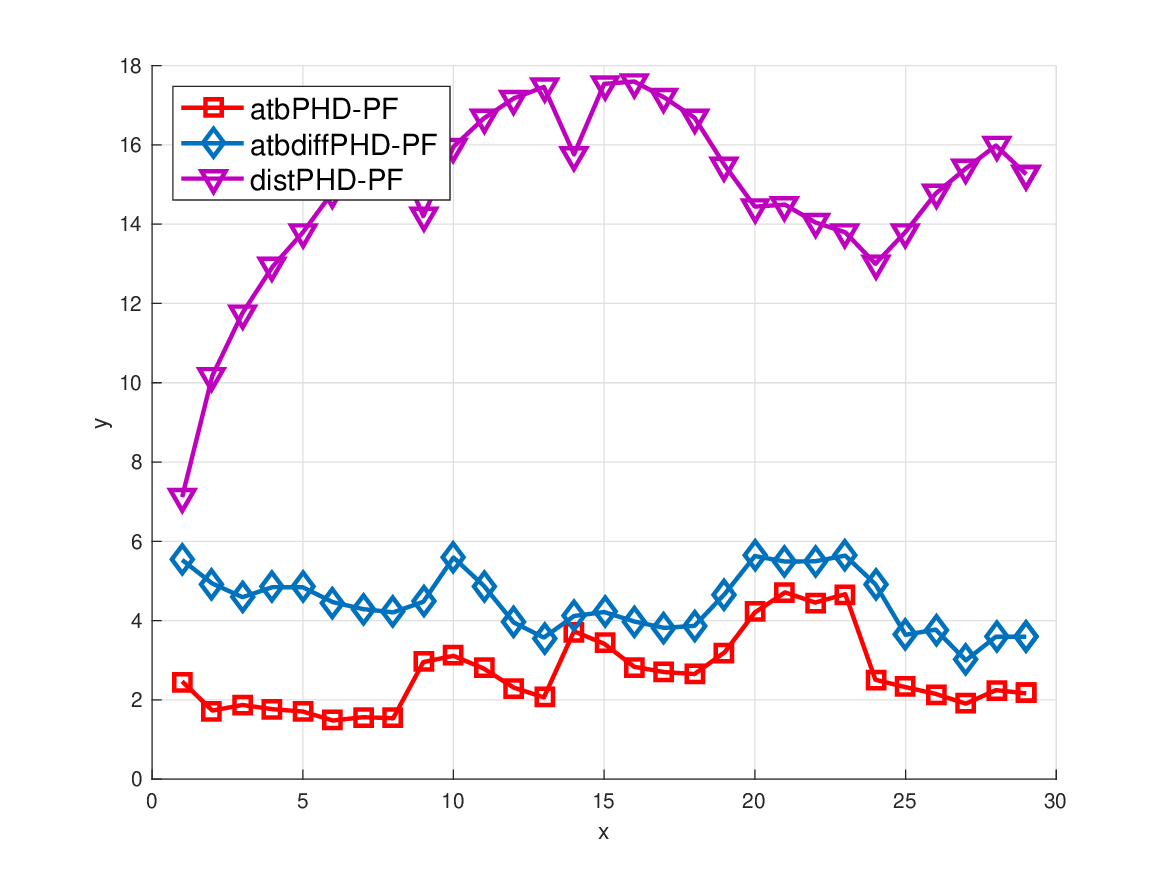}
\qquad
\psfrag{y}[bc][cc][0.9]{\footnotesize $N_\text{est}$}
\includegraphics[width=.45\textwidth]{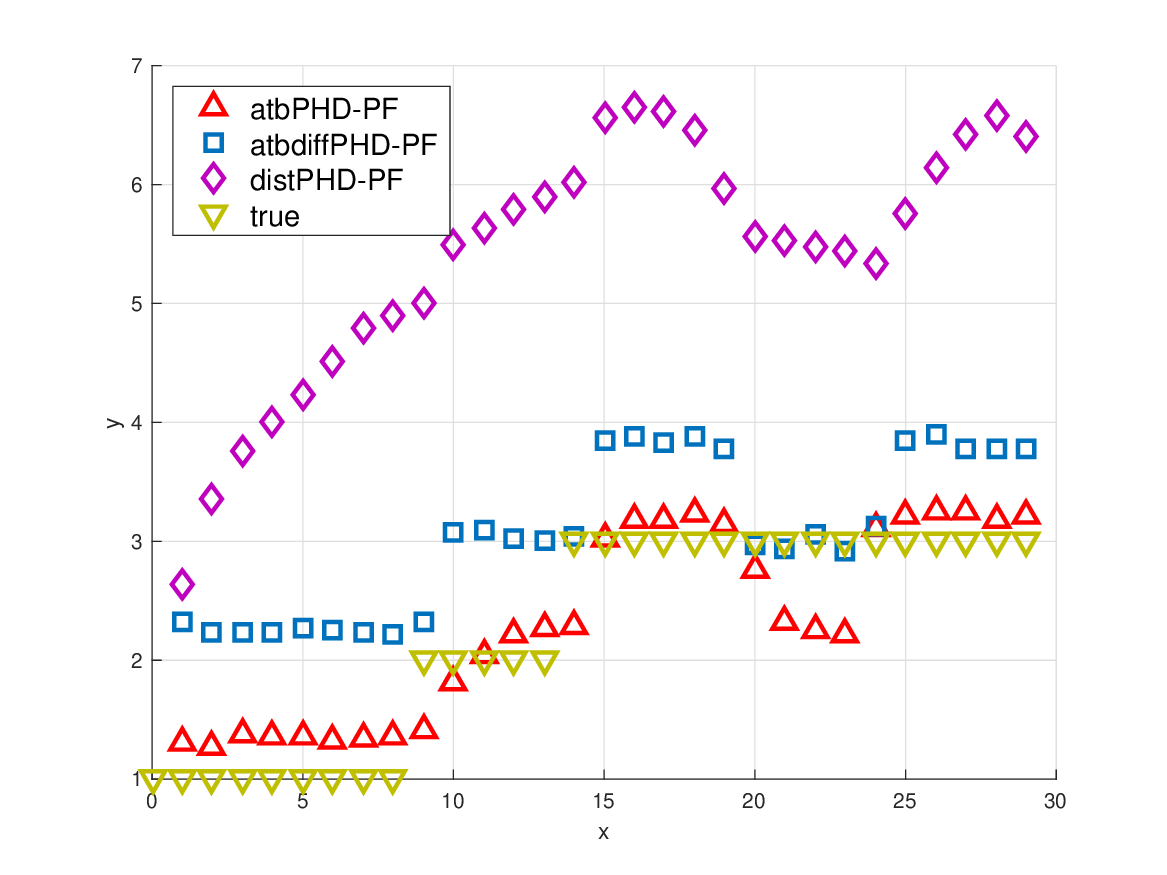}
\caption{Simulation II: Results for Gaussian noise and clutter rate $\lambda_\text{FA}=0.3$. The left part of the figure shows the squared and scaled \acl{ospa} (\acs{ospa}) metric for each algorithm compared to the \acl{dpcrlb} (\acs{dpcrlb}), while the right part compares the estimated to the true number of targets.}
\label{fig:simresults1}
\end{figure*}

\subsection{Simulation II: Results}
In the second simulation, we evaluate the performance of the MS-PPHDF, the D-PPHDF and the DDF-PPHDF under a higher clutter rate of $0.3$. The remaining parameters are chosen as in the previous simulation. The simulation results are shown in Figure \ref{fig:simresults1}. The top part considers zero-mean Gaussian measurement noise with a per-component variance of $\sigma_r^2 = 0.1$ while the bottom part shows the results for $\sigma_r^2 = 0.3$.

While the higher clutter rate causes an increase in the OSPA value of all algorithms, the \ac{ms-pphdf} is still able to correctly estimate the number of targets (except for the crossing period $20 \leq i \leq 24$) in both cases. When taking the next lower integer of the estimate, the \ac{phd-dpf} also yields acceptable results for $\sigma_r^2=0.1$. For $\sigma_r^2=0.3$ the number of targets is overestimated by 1 for $1\leq i \leq 15$, causing a stronger degradation of the scaled and squared OPSA value in this interval.

Apparently, the \ac{ddf-pphdf} is not able to cope with a clutter rate of $0.3$ as the number of targets is largely overestimated. Hence, no accurate target tracking is possible.

\subsection{Simulation III: Results}
In the third simulation, we evaluate the robustness of the MS-PPHDF, the D-PPHDF and the DDF-PPHDF in the face of $\varepsilon$-contaminated noise and different clutter rates. We consider a per-component variance of the measurement noise of $\sigma_r^2=0.1$ and $0.3$, an average number of clutter of $\lambda_\text{FA}=0.1$ and $0.3$, as well as a contamination of $10 \%$ and $30 \%$. The remaining parameters are chosen as before. The simulation results for clutter rates $\lambda_\text{FA}=0.1$ and $0.3$ are given in Figures~\ref{fig:simresults2} and \ref{fig:simresults3}, respectively. The top half of each figure considers $\sigma_r^2=0.1$ while the bottom half pertains to $\sigma_r^2=0.3$. Rows 1 and 3 deal with a noise contamination of $10~\%$, rows 2 and 4 show the results for $30~\%$.
 
Let us look at the case of $\lambda_\text{FA}=0.1$ first. We observe that the centralized \ac{ms-pphdf} is still the best performing algorithm, being largely unaffected by higher noise variance and outliers. The \ac{phd-dpf} is a close second, being primarily affected by the higher clutter rate and the higher noise variance. It shows only a slight additional performance degradation when increasing the noise contamination to $30~\%$. Hence, it can be said that both algorithms are robust against outliers and can handle a fraction of at least $10~\%$ in the given scenario. The \ac{ms-pphdf} can also cope with $\lambda_\text{FA}=0.3$ while the target position estimates of the \ac{phd-dpf} might be too imprecise in this case, depending on the problem at hand. 
The \ac{ddf-pphdf}, in contrast, is more severely affected by outliers. Both the OSPA value and the estimated number of targets increase with the introduction of noise contamination. When the number of targets remains constant and no target crossing takes place, i.e. for $i < 9$ and $i > 24$,  the number of targets is only slightly overestimated. However, when targets two and three enter the scene, i.e. for $10 \leq i < 20$, the estimate is inaccurate, which imposes a penalty on the scaled and squared OPSA metric. Hence, the \ac{ddf-pphdf} is not a robust algorithm for the considered tracking scenario.

In the case of $\lambda_\text{FA}=0.3$, the \ac{ddf-pphdf}, again, breaks down completely. The \ac{ms-pphdf}, however, is still able to give accurate results with only slight deviations from the true number of targets and a small OSPA value. Unfortunately, the combination of noise contamination and more clutter is too much for the \ac{phd-dpf} to handle. It overestimates the number of targets by one to two, causing the OSPA value to rise as well. 

In summary, the proposed \ac{ms-pphdf} and \ac{phd-dpf} are---to a certain extent---robust against outliers of the $\varepsilon$-contamination kind. This property is due to the employed two-way communication scheme, which vets measurements as well as intermediate target position estimates against the entire network or the neighborhood of each node. The alternative \ac{ddf-pphdf}, however, breaks down in the face of outliers.

In \cite{hou2017robust,leonard2017robusta,leonard2018robust}, we successfully proposed to use robust estimators to robustify sequential detectors for distributed sensor networks. We applied the same concept to the distributed \ac{phd-dpf}. However, no further performance improvement could be gained here since the twofold neighborhood averaging already exhausted the power of neighborhood communication. 
\begin{figure*}[!t]
\psfrag{atbPHD-PF}[cl][cl][0.8]{\footnotesize MS-PPHDF}
\psfrag{atbdiffPHD-PF}[cl][cl][0.8]{\footnotesize D-PPHDF}
\psfrag{distPHD-PF}[cl][cl][0.8]{\footnotesize DDF-PPHDF}
\psfrag{PCRLB}[cl][cl][0.8]{\footnotesize DPCRLB}
\psfrag{true}[cl][cl][0.8]{\footnotesize true number}
\psfrag{y}[bc][cc][0.9]{\footnotesize $N_\text{tgt}\cdot\left(\bar{d}_2^{(2)}\right)^2$ and $B_{i,\text{dist}}$}
\psfrag{x}[tc][tc][0.9]{\footnotesize Time steps $i$ (seconds)}
\psfrag{0}[Bc][Bc][0.7]{\footnotesize 0}
\psfrag{5}[cc][cc][0.7]{\footnotesize 5}
\psfrag{15}[cc][cc][0.7]{\footnotesize 15}
\psfrag{25}[cc][cc][0.7]{\footnotesize 25}
\psfrag{10}[cc][cc][0.7]{\footnotesize 10}
\psfrag{20}[cc][cc][0.7]{\footnotesize 20}
\psfrag{30}[cc][cc][0.7]{\footnotesize 30}
\psfrag{35}[cc][cc][0.7]{\footnotesize 35}
\psfrag{0.2}[cc][cc][0.7]{\footnotesize 0.2}
\psfrag{0.4}[cc][cc][0.7]{\footnotesize 0.4}
\psfrag{0.6}[cc][cc][0.7]{\footnotesize 0.6}
\psfrag{0.8}[cc][cc][0.7]{\footnotesize 0.8}
\psfrag{1}[cc][cc][0.7]{\footnotesize 1}
\psfrag{1.2}[cc][cc][0.7]{\footnotesize 1.2}
\psfrag{1.4}[cc][cc][0.7]{\footnotesize 1.4}
\psfrag{1.6}[cc][cc][0.7]{\footnotesize 1.6}
\psfrag{1.8}[cc][cc][0.7]{\footnotesize 1.8}
\psfrag{2}[cc][cc][0.7]{\footnotesize 2}
\centering
\begin{tabular}[b]{c}
	$\sigma_r^2=0.1$\\ 
	$\varepsilon=0.1$\\
	\\\\\\\\\\\\
\end{tabular}
\includegraphics[width=.44\textwidth]{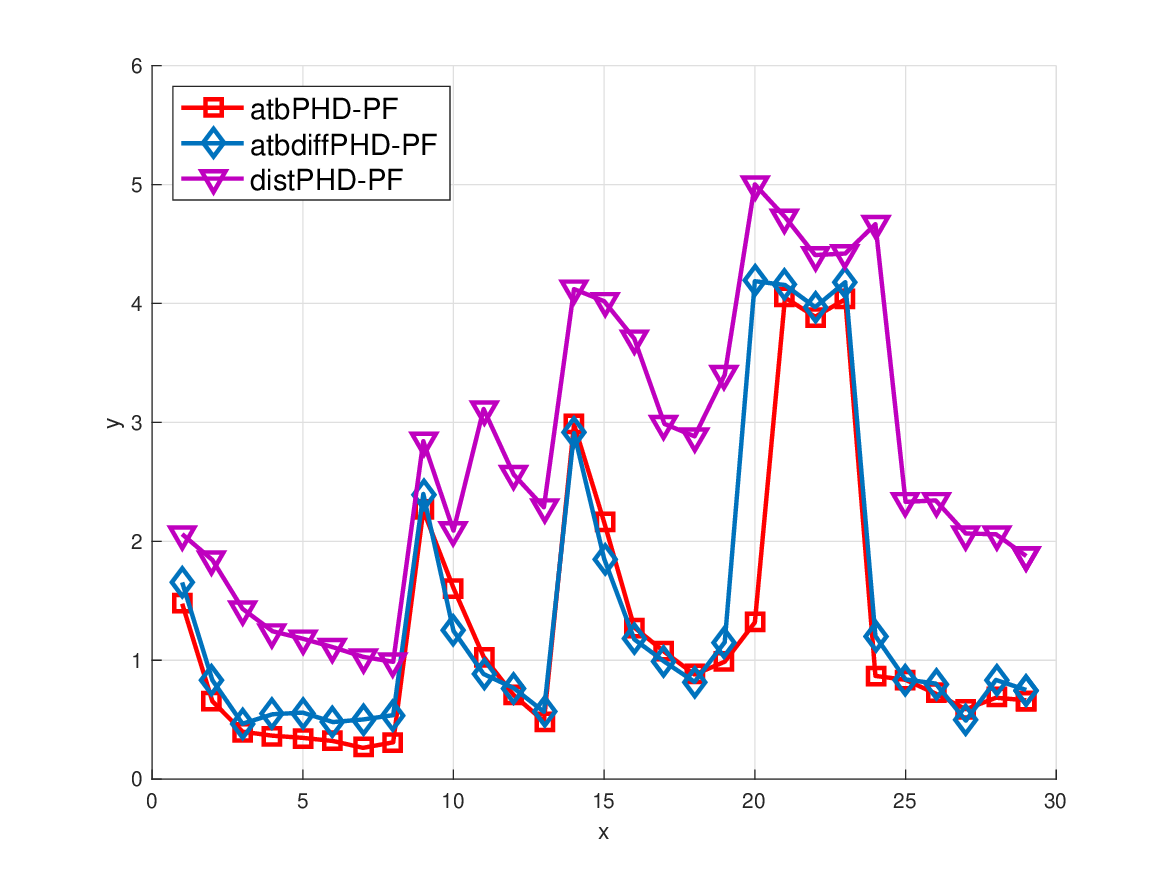}
\psfrag{y}[bc][cc][0.9]{\footnotesize $N_\text{est}$}
\includegraphics[width=.44\textwidth]{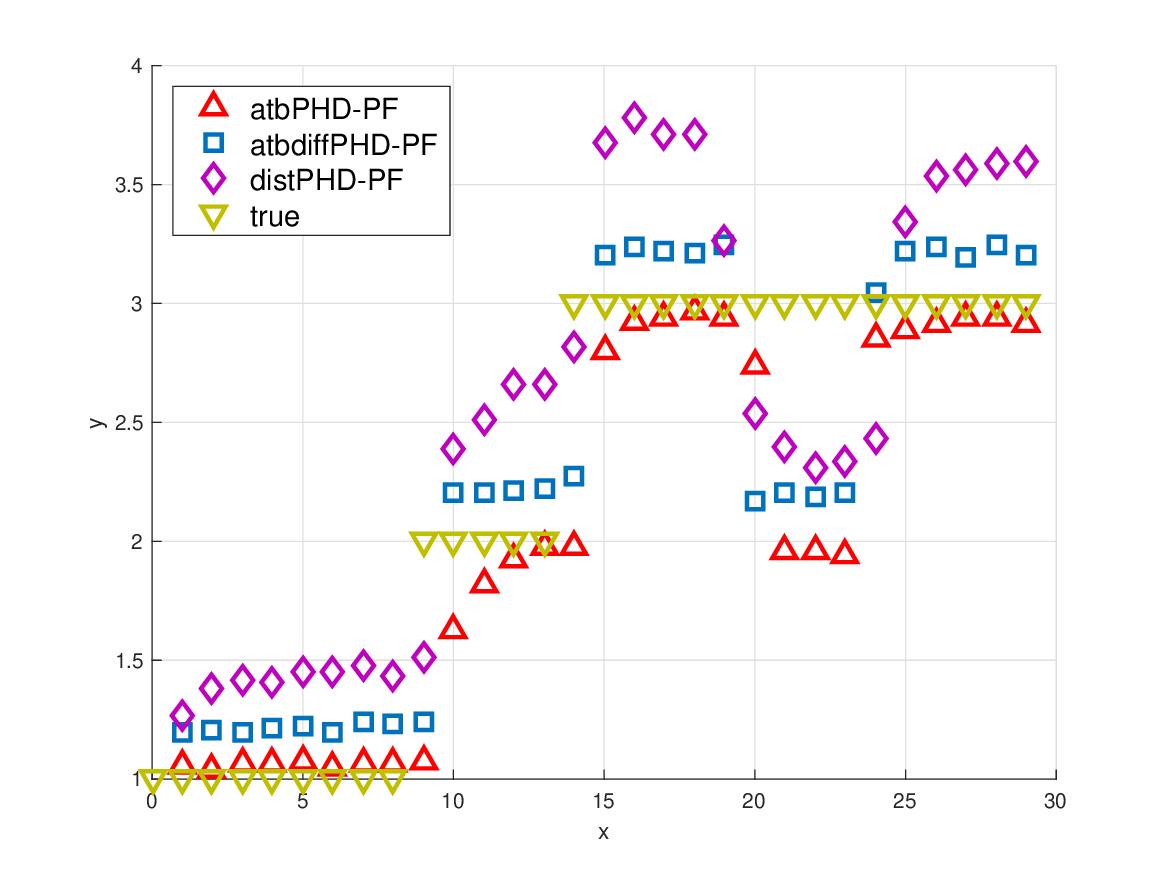}
\\
\begin{tabular}[b]{c}
	$\sigma_r^2=0.1$\\ 
	$\varepsilon=0.3$\\
	\\\\\\\\\\\\
\end{tabular}
\psfrag{y}[bc][cc][0.9]{\footnotesize $N_\text{tgt}\cdot\left(\bar{d}_2^{(2)}\right)^2$ and $B_{i,\text{dist}}$}
\includegraphics[width=.44\textwidth]{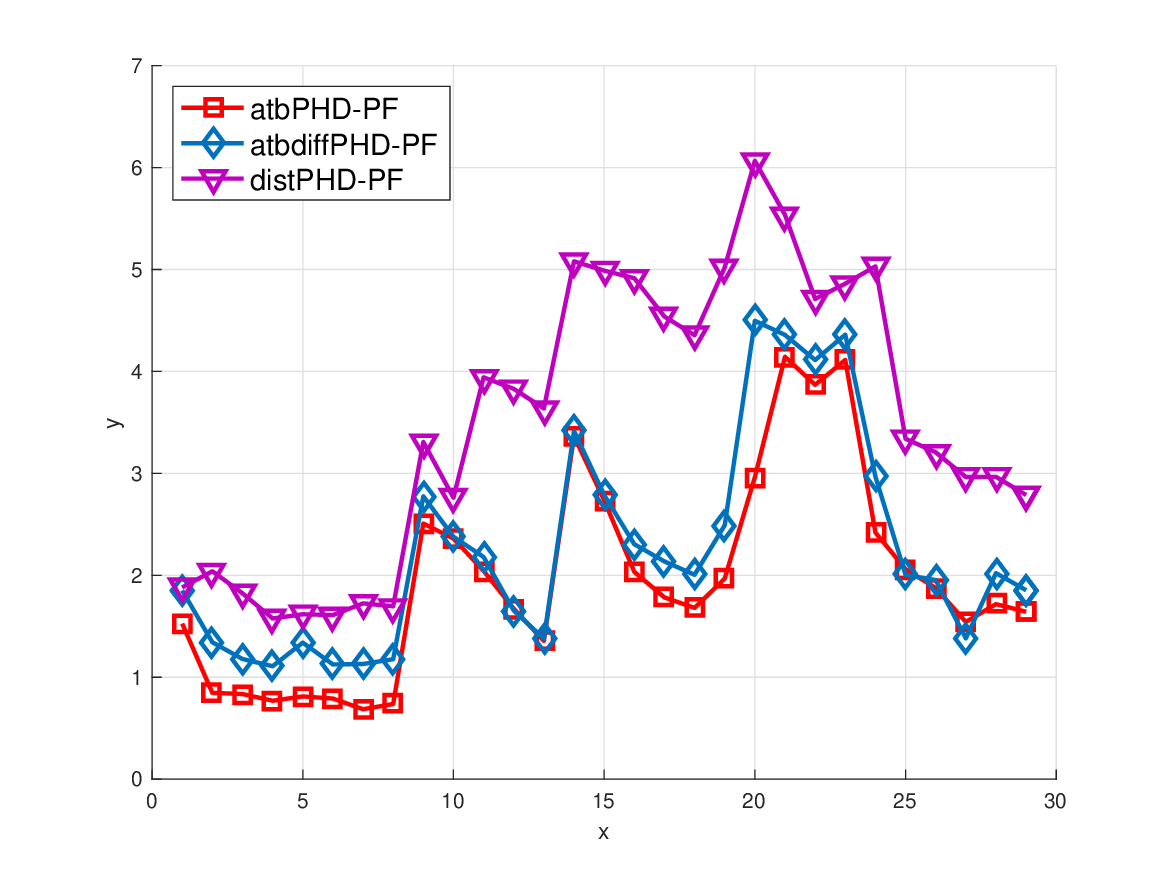}
\psfrag{y}[bc][cc][0.9]{\footnotesize $N_\text{est}$}
\includegraphics[width=.44\textwidth]{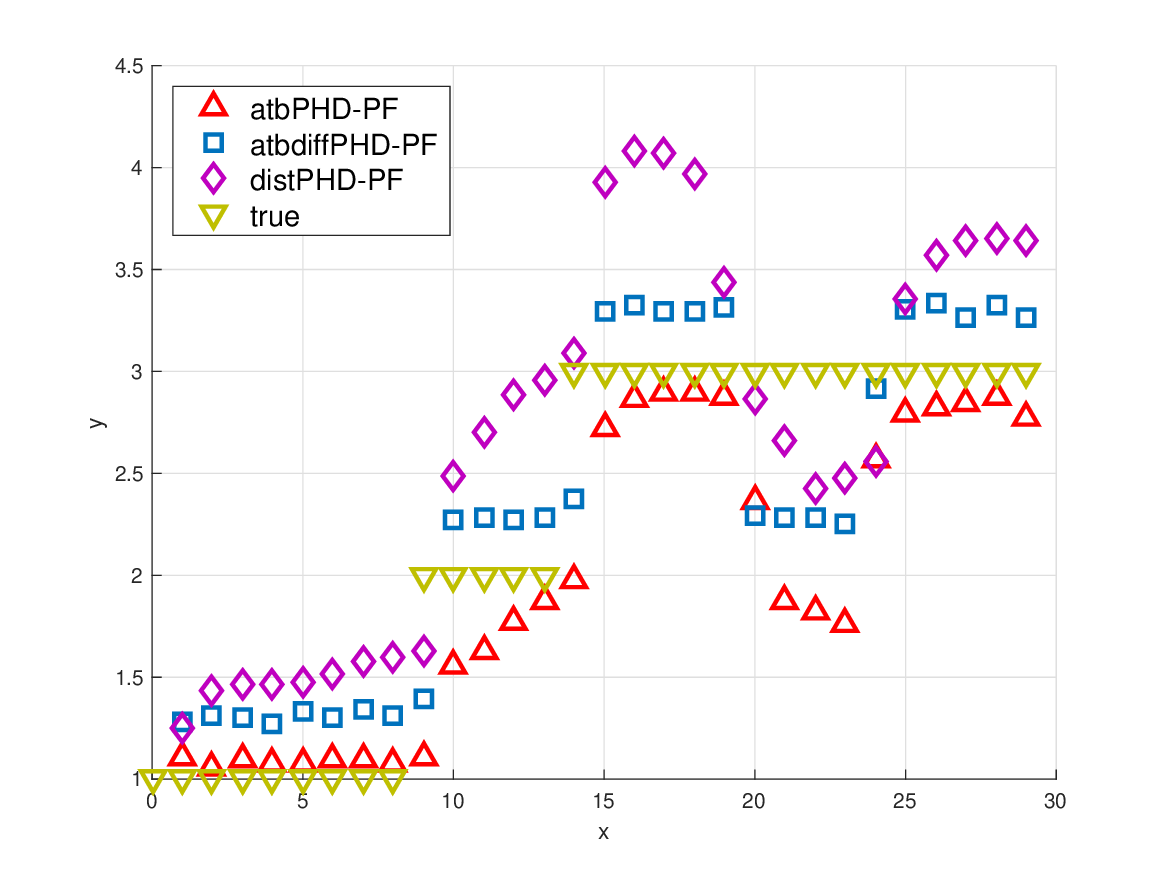}
\\
\begin{tabular}[b]{c}
	$\sigma_r^2=0.3$\\ 
	$\varepsilon=0.1$\\
	\\\\\\\\\\\\
\end{tabular}
\psfrag{y}[bc][cc][0.9]{\footnotesize $N_\text{tgt}\cdot\left(\bar{d}_2^{(2)}\right)^2$ and $B_{i,\text{dist}}$}
\includegraphics[width=.44\textwidth]{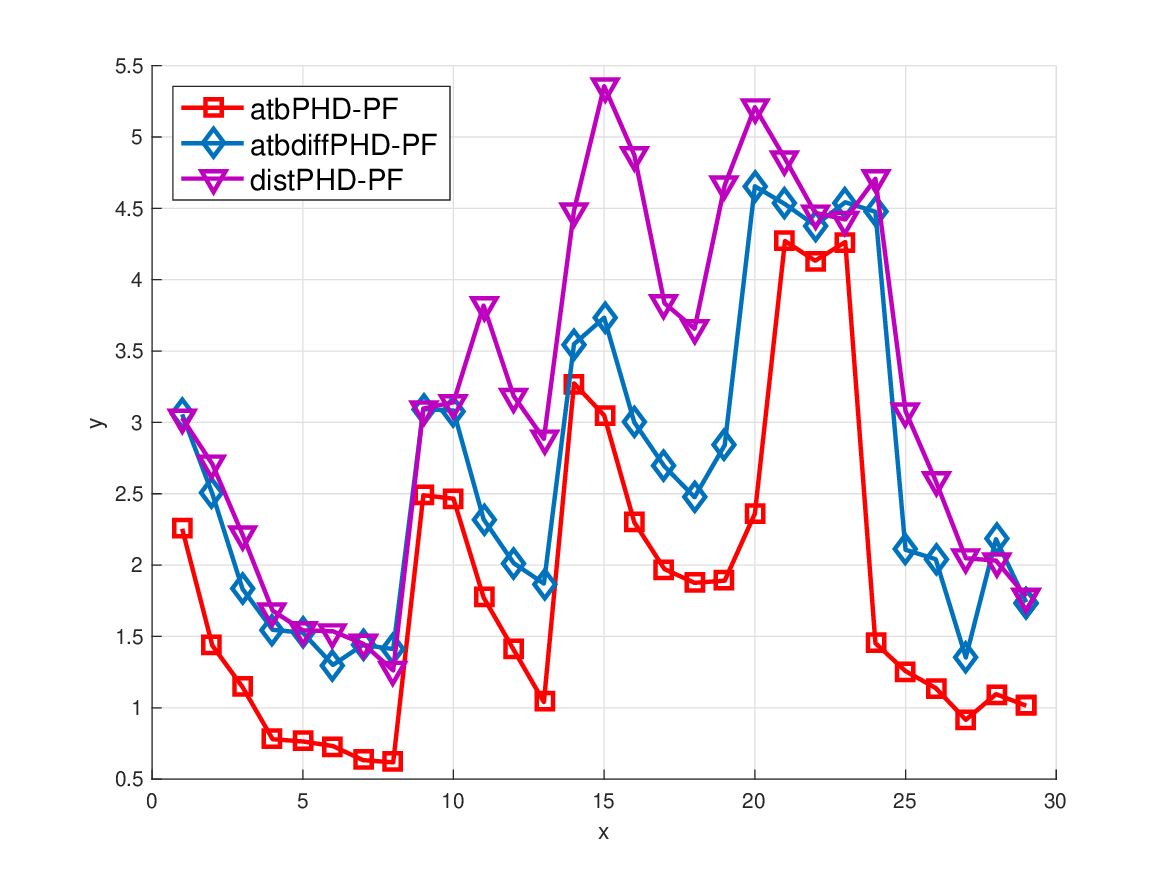}
\psfrag{y}[bc][cc][0.9]{\footnotesize $N_\text{est}$}
\includegraphics[width=.44\textwidth]{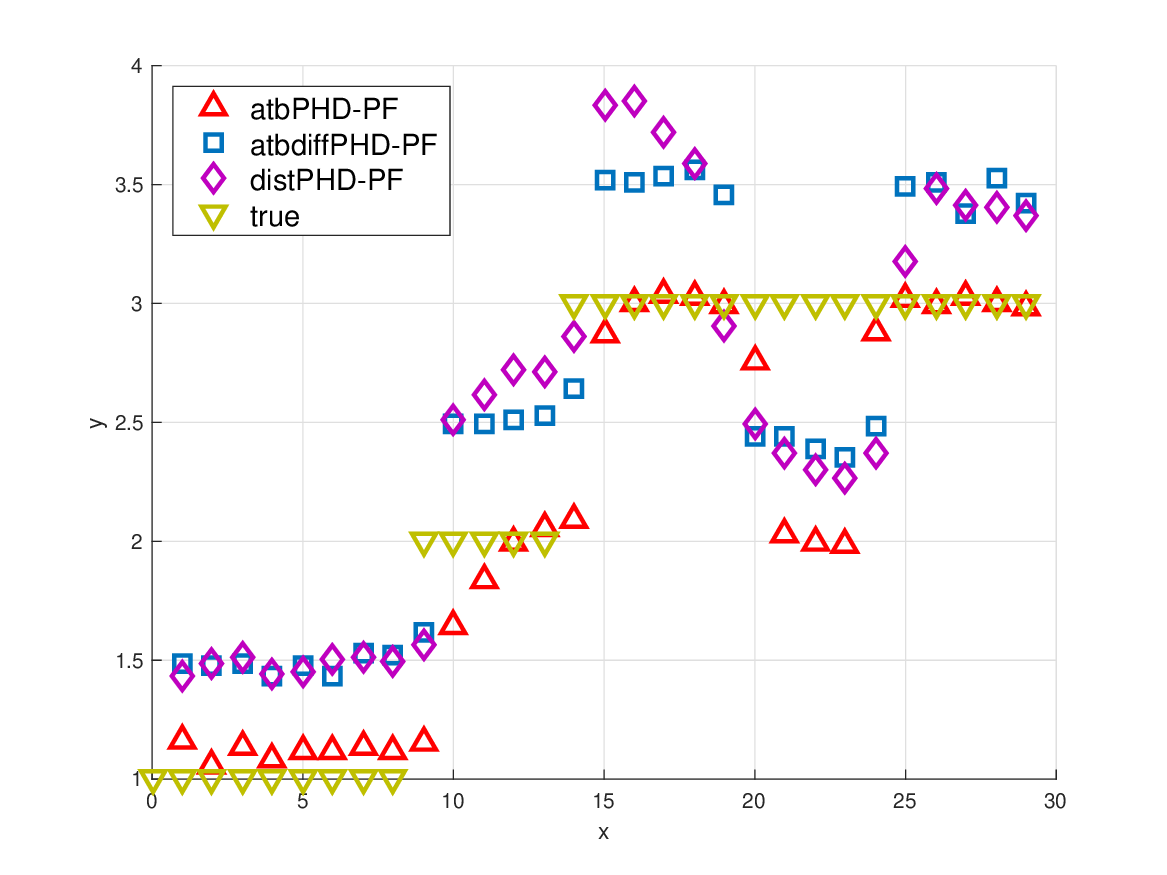}
\\
\begin{tabular}[b]{c}
	$\sigma_r^2=0.3$\\ 
	$\varepsilon=0.3$\\
	\\\\\\\\\\\\
\end{tabular}
\psfrag{y}[bc][cc][0.9]{\footnotesize $N_\text{tgt}\cdot\left(\bar{d}_2^{(2)}\right)^2$ and $B_{i,\text{dist}}$}
\includegraphics[width=.44\textwidth]{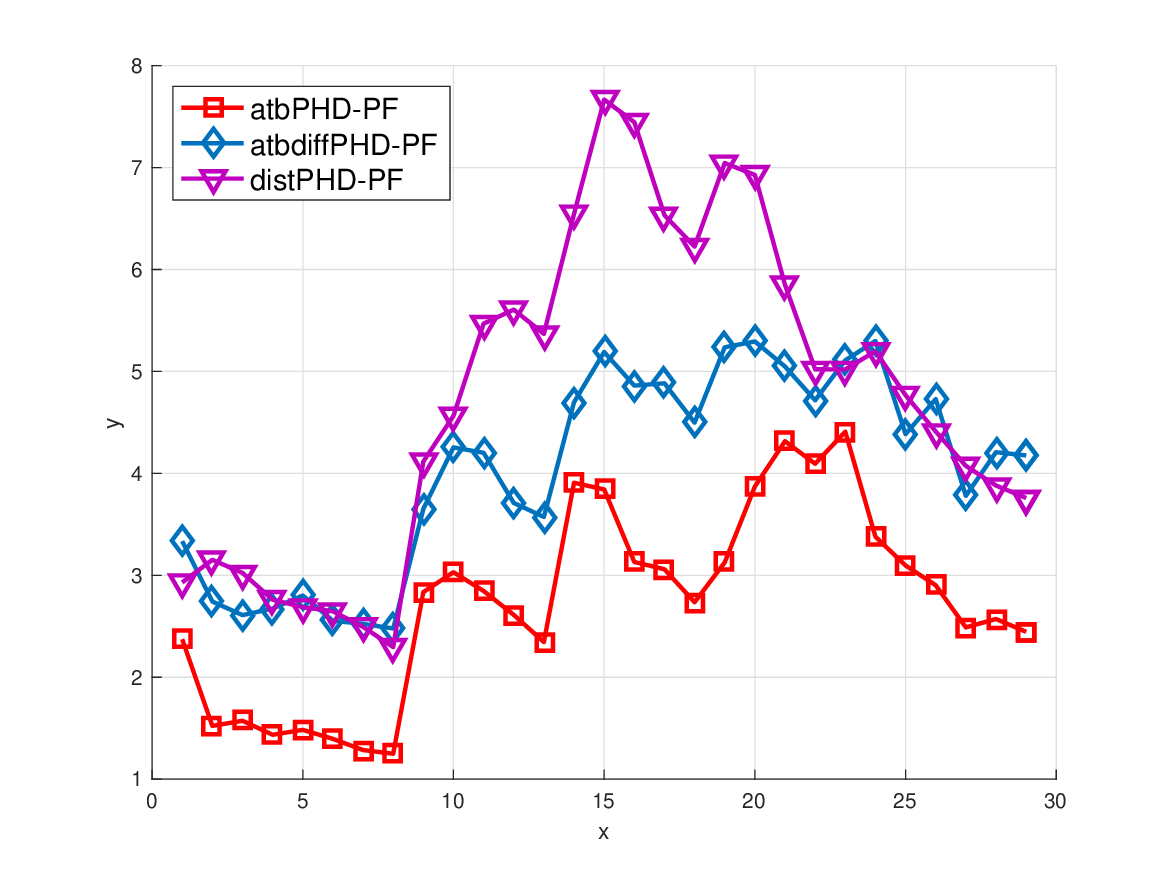}
\psfrag{y}[bc][cc][0.9]{\footnotesize $N_\text{est}$}
\includegraphics[width=.44\textwidth]{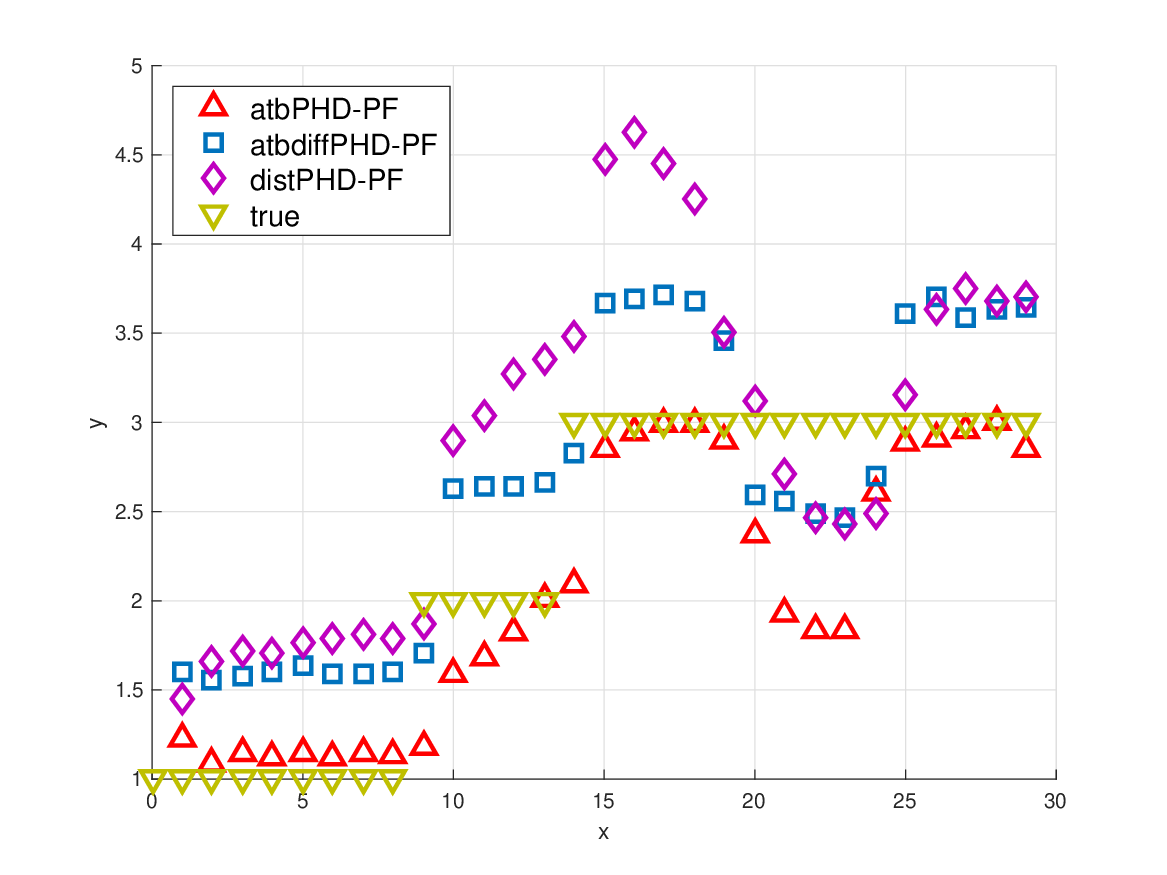}
\caption{Simulation III: Results for $\varepsilon$-contaminated noise and clutter rate $\lambda_\text{FA}=0.1$. The two upper rows consider $\sigma_r^2=0.1$, the lower ones show the results for $\sigma_r^2=0.3$. Rows 1 and 3 consider $\varepsilon=0.1$, rows 2 and 4 show the results for $\varepsilon=0.3$.}
\label{fig:simresults3}
\end{figure*}

\begin{figure*}[!t]
\psfrag{atbPHD-PF}[cl][cl][0.8]{\footnotesize MS-PPHDF}
\psfrag{atbdiffPHD-PF}[cl][cl][0.8]{\footnotesize D-PPHDF}
\psfrag{distPHD-PF}[cl][cl][0.8]{\footnotesize DDF-PPHDF}
\psfrag{PCRLB}[cl][cl][0.8]{\footnotesize DPCRLB}
\psfrag{true}[cl][cl][0.8]{\footnotesize true number}
\psfrag{y}[bc][cc][0.9]{\footnotesize $N_\text{tgt}\cdot\left(\bar{d}_2^{(2)}\right)^2$ and $B_{i,\text{dist}}$}
\psfrag{x}[tc][tc][0.9]{\footnotesize Time steps $i$ (seconds)}
\psfrag{0}[Bc][Bc][0.7]{\footnotesize 0}
\psfrag{5}[cc][cc][0.7]{\footnotesize 5}
\psfrag{15}[cc][cc][0.7]{\footnotesize 15}
\psfrag{25}[cc][cc][0.7]{\footnotesize 25}
\psfrag{10}[cc][cc][0.7]{\footnotesize 10}
\psfrag{20}[cc][cc][0.7]{\footnotesize 20}
\psfrag{30}[cc][cc][0.7]{\footnotesize 30}
\psfrag{35}[cc][cc][0.7]{\footnotesize 35}
\psfrag{0.2}[cc][cc][0.7]{\footnotesize 0.2}
\psfrag{0.4}[cc][cc][0.7]{\footnotesize 0.4}
\psfrag{0.6}[cc][cc][0.7]{\footnotesize 0.6}
\psfrag{0.8}[cc][cc][0.7]{\footnotesize 0.8}
\psfrag{1}[cc][cc][0.7]{\footnotesize 1}
\psfrag{1.2}[cc][cc][0.7]{\footnotesize 1.2}
\psfrag{1.4}[cc][cc][0.7]{\footnotesize 1.4}
\psfrag{1.6}[cc][cc][0.7]{\footnotesize 1.6}
\psfrag{1.8}[cc][cc][0.7]{\footnotesize 1.8}
\psfrag{2}[cc][cc][0.7]{\footnotesize 2}
\centering
\begin{tabular}[b]{c}
	$\sigma_r^2=0.1$\\ 
	$\varepsilon=0.1$\\
	\\\\\\\\\\\\
\end{tabular}
\includegraphics[width=.44\textwidth]{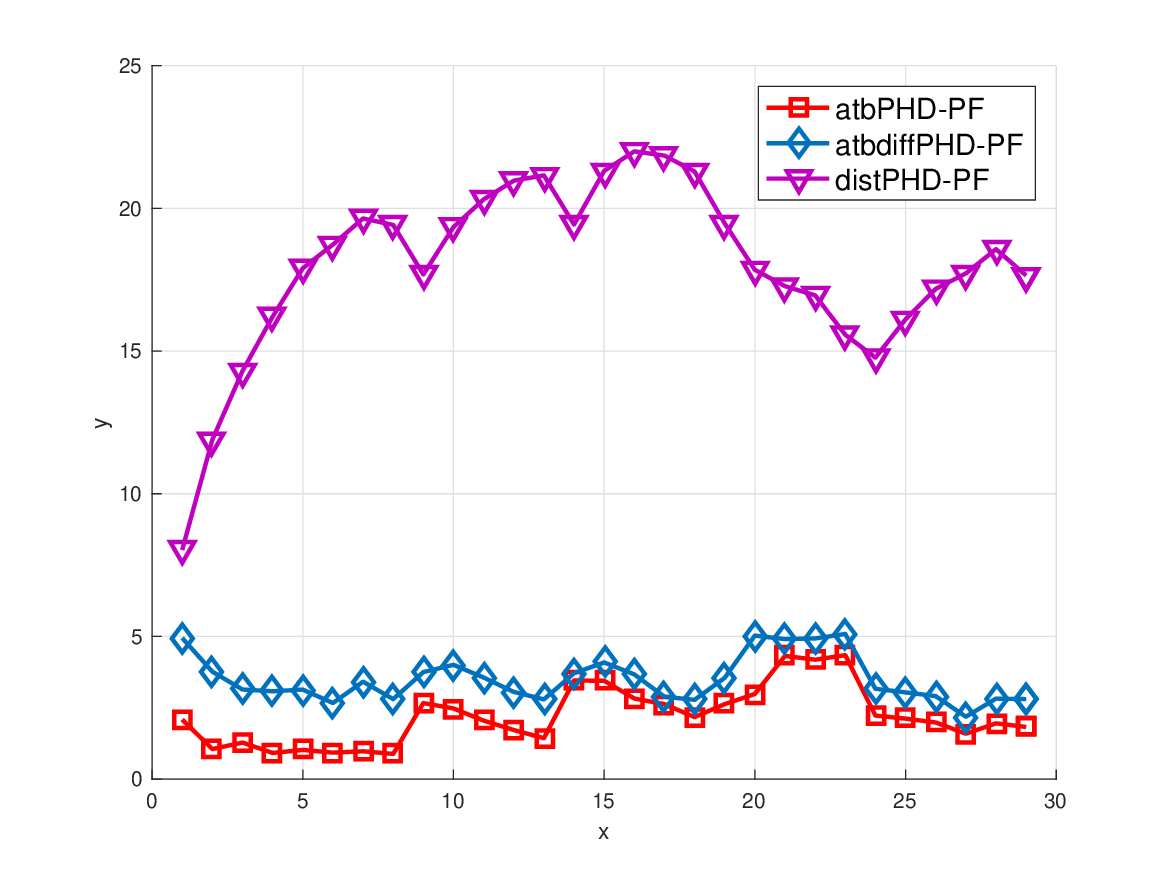}
\psfrag{y}[bc][cc][0.9]{\footnotesize $N_\text{est}$}
\includegraphics[width=.44\textwidth]{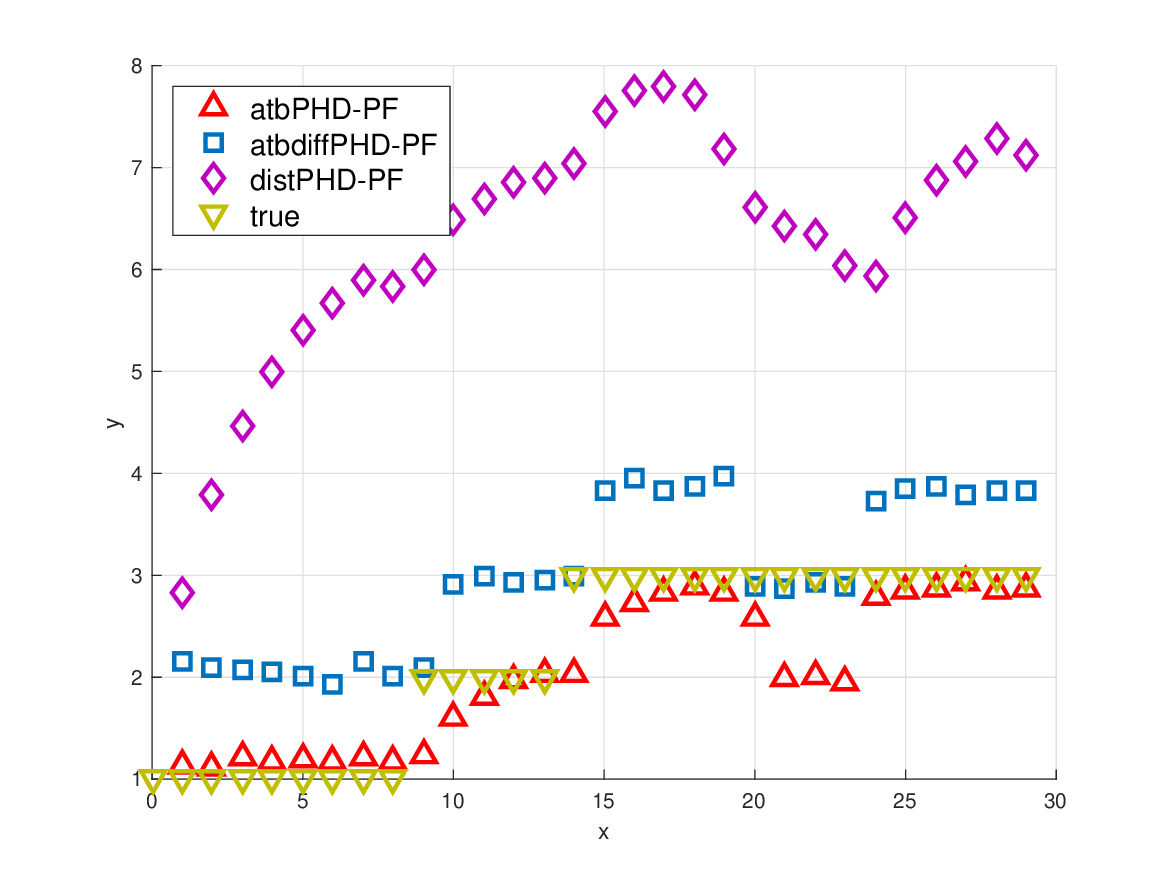}
\\
\begin{tabular}[b]{c}
	$\sigma_r^2=0.1$\\ 
	$\varepsilon=0.3$\\
	\\\\\\\\\\\\
\end{tabular}
\psfrag{y}[bc][cc][0.9]{\footnotesize $N_\text{tgt}\cdot\left(\bar{d}_2^{(2)}\right)^2$ and $B_{i,\text{dist}}$}
\includegraphics[width=.44\textwidth]{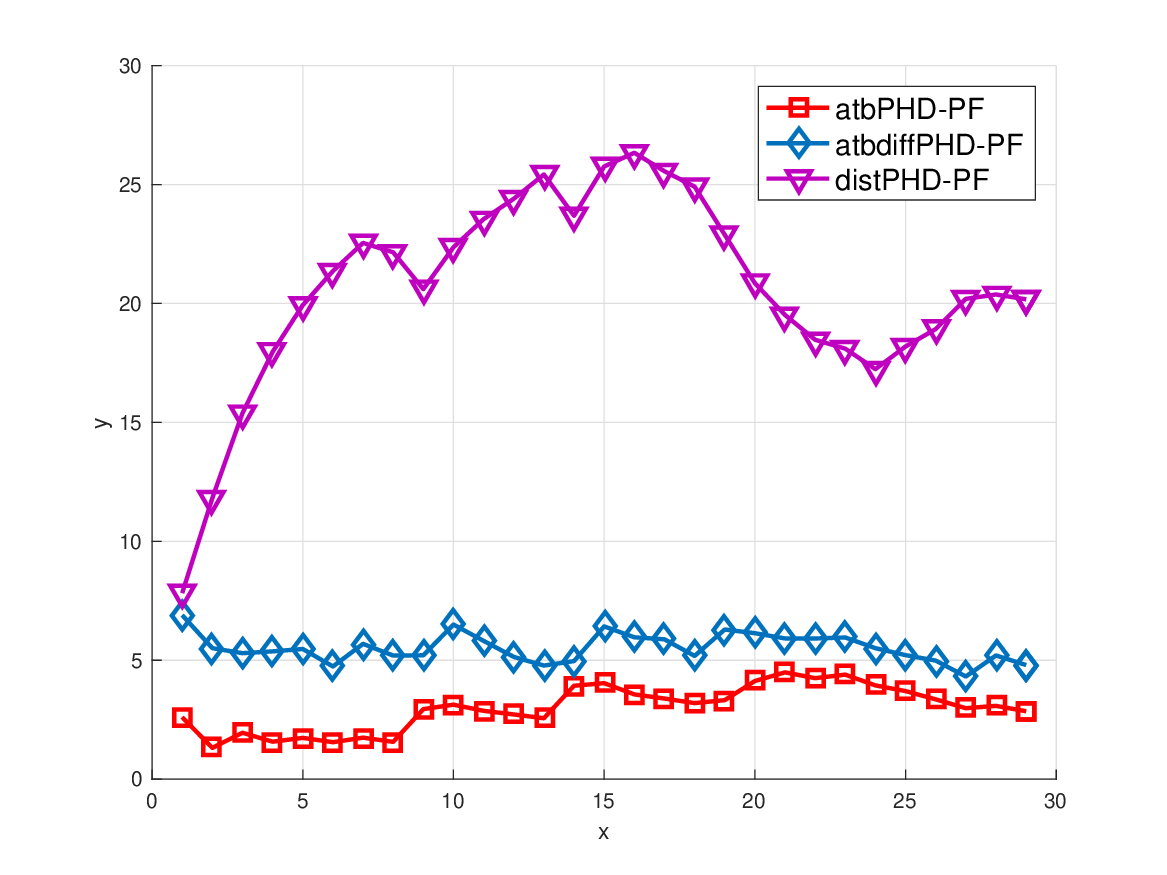}
\psfrag{y}[bc][cc][0.9]{\footnotesize $N_\text{est}$}
\includegraphics[width=.44\textwidth]{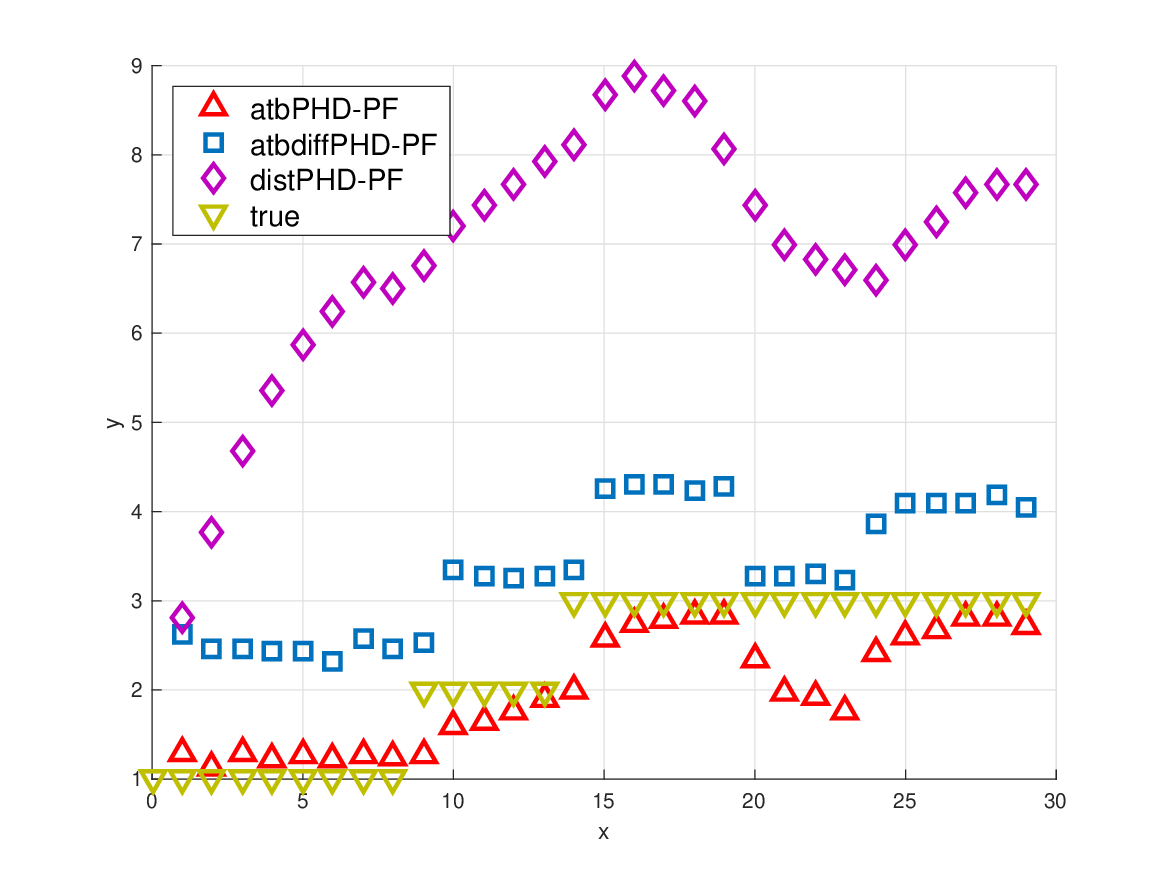}
\\
\begin{tabular}[b]{c}
	$\sigma_r^2=0.3$\\ 
	$\varepsilon=0.1$\\
	\\\\\\\\\\\\
\end{tabular}
\psfrag{y}[bc][cc][0.9]{\footnotesize $N_\text{tgt}\cdot\left(\bar{d}_2^{(2)}\right)^2$ and $B_{i,\text{dist}}$}
\includegraphics[width=.44\textwidth]{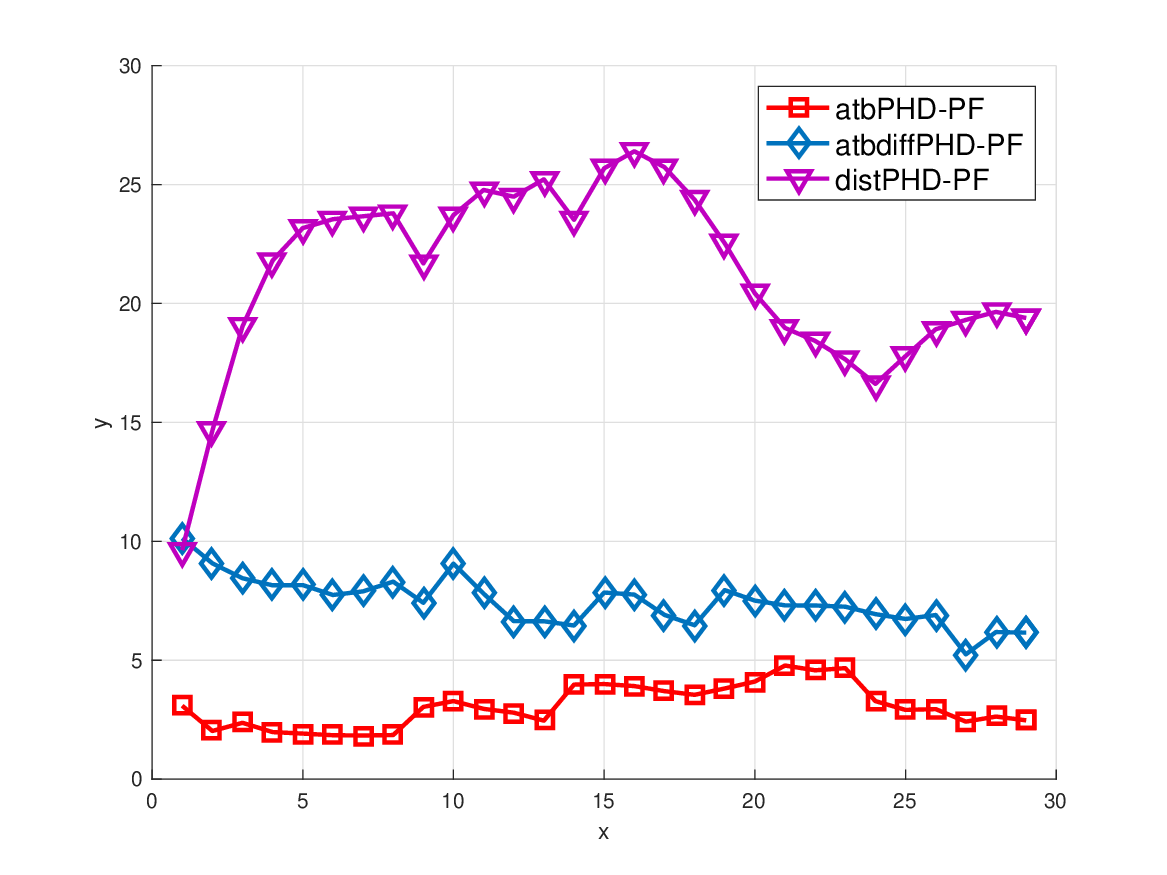}
\psfrag{y}[bc][cc][0.9]{\footnotesize $N_\text{est}$}
\includegraphics[width=.44\textwidth]{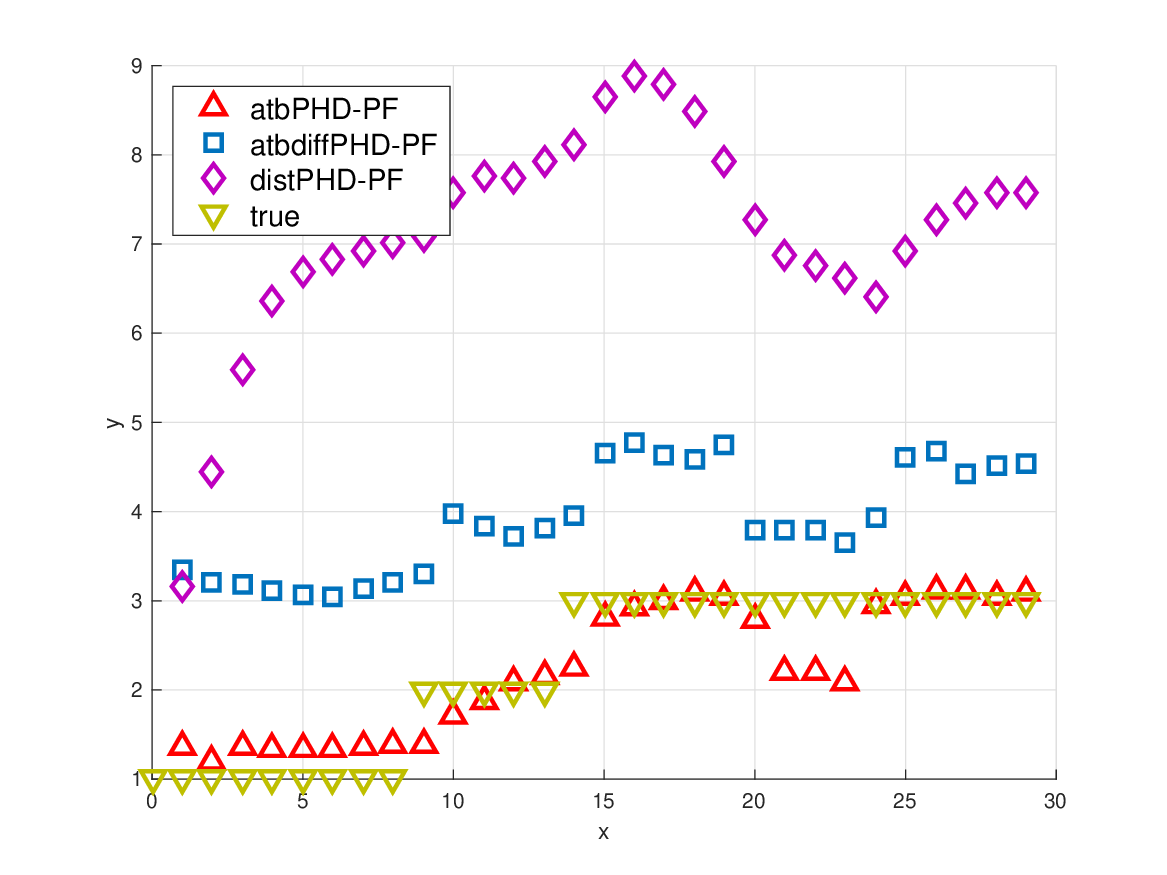}
\\
\begin{tabular}[b]{c}
	$\sigma_r^2=0.3$\\ 
	$\varepsilon=0.3$\\
	\\\\\\\\\\\\
\end{tabular}
\psfrag{y}[bc][cc][0.9]{\footnotesize $N_\text{tgt}\cdot\left(\bar{d}_2^{(2)}\right)^2$ and $B_{i,\text{dist}}$}
\includegraphics[width=.44\textwidth]{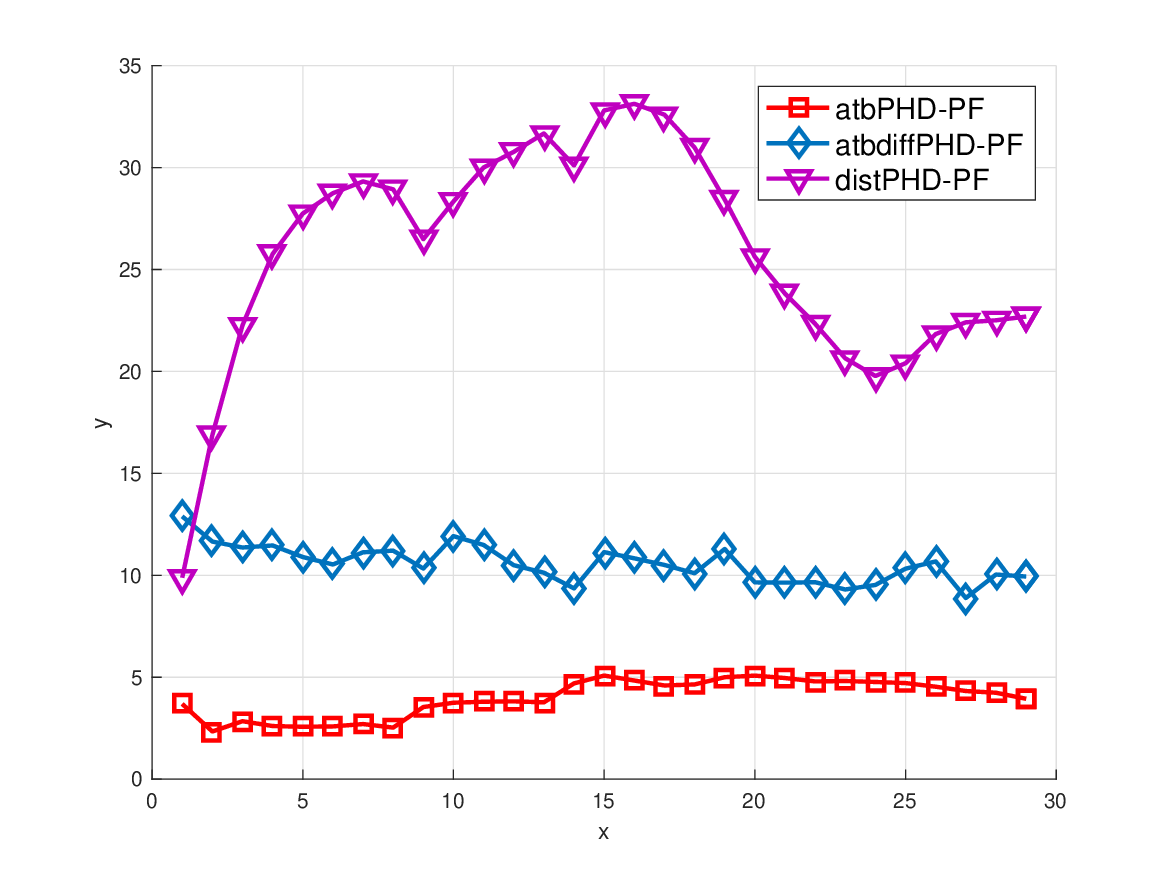}
\psfrag{y}[bc][cc][0.9]{\footnotesize $N_\text{est}$}
\includegraphics[width=.44\textwidth]{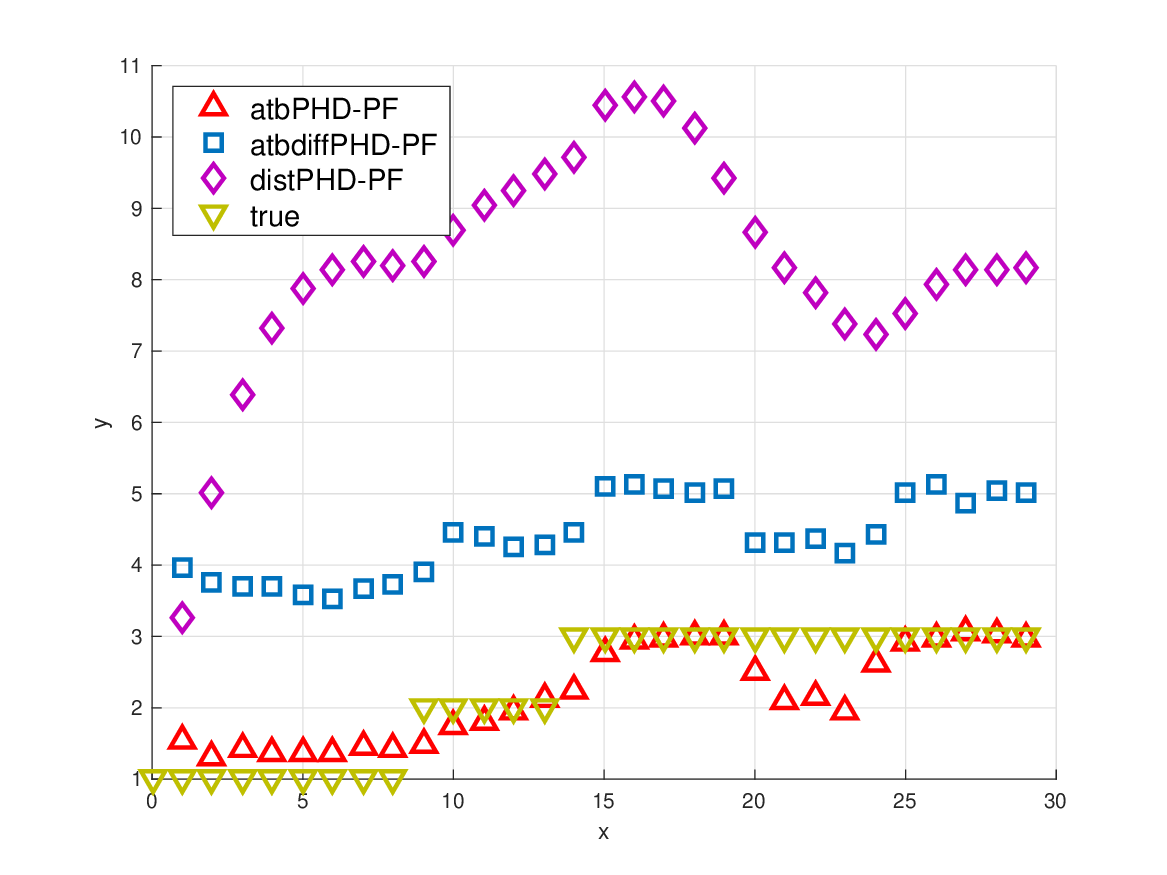}
\caption{Simulation III: Results for $\varepsilon$-contaminated noise and clutter rate $\lambda_\text{FA}=0.3$. The two upper rows consider $\sigma_r^2=0.1$, the lower ones show the results for $\sigma_r^2=0.3$. Rows 1 and 3 consider $\varepsilon=0.1$, rows 2 and 4 show the results for $\varepsilon=0.3$.}
\label{fig:simresults3}
\end{figure*}

\section{Conclusion}
In this work, we developed a distributed as well as a centralized \ac{phd-pf} for \ac{mtt} in sensor networks. We, furthermore, came up with a distributed version of the \ac{pcrlb} that served as a benchmark in the performance evaluation. Our simulation results showed that the distributed \ac{phd-dpf} is faster in correctly tracking new targets than the centralized \ac{ms-pphdf} and performs only slightly worse when the number of targets stays constant. In addition, it delivers accurate tracking results as long as the targets are far enough apart so that their corresponding measurement clouds are separable. Our approach outperforms the existing \ac{ddf-pphdf} at the cost of additional communication between sensor nodes. Moreover, the proposed trackers are inherently robust against outliers and the centralized \ac{ms-pphdf} is even able to handle higher clutter rates. The existing \ac{ddf-pphdf}, in contrast, is neither robust nor able to cope with more than $10~\%$ clutter.

\section*{Acknowledgment}
The authors would like to thank Dr. Paolo Braca from the NATO Science \& Technology Organization, Centre for Maritime Research and Experimentation in La Spezia, Italy, for his valuable comments.

\ifCLASSOPTIONcaptionsoff
  \newpage
\fi



%
\bibliographystyle{IEEEtran}
\bibliography{references.bib}

\end{document}